\documentclass[aps,pra,twocolumn,floats,floatfix,showpacs,superscriptaddress]{revtex4-1}

\usepackage{bbm}
\usepackage{amssymb}
\usepackage{ifpdf}
\usepackage{color}
\usepackage{graphicx,epsfig}

\usepackage[apple mac]{inputenc}

\usepackage{amsmath}
\usepackage{hyperref}

\renewcommand{\thefigure}{\textbf{\arabic{figure}}}

\newcommand{\nocontentsline}[3]{}
\newcommand{\tocless}[2]{\vspace{5mm}\bgroup\let\addcontentsline=\nocontentsline#1{#2}\egroup}

\newcommand{\ket}[1]{\ensuremath{|#1\rangle}}
\newcommand{\bra}[1]{\ensuremath{\langle#1|}}
\newcommand{\qone}{\ensuremath{\uparrow}}
\newcommand{\qzero}{\ensuremath{\downarrow}}
\newcommand{\aone}{\ensuremath{1}}
\newcommand{\azero}{\ensuremath{0}}

\newcommand{\be}{\begin{equation}}
\newcommand{\ee}{\end{equation}}

\newcommand{\si}[1]{(see Appendix~#1)}

\newcommand{\1}{{\mathbbm{1}}}

\newcommand{\uibke}{Institut f\"ur Experimentalphysik, Universit\"at Innsbruck, Technikerstrasse 25, 6020 Innsbruck, Austria}
\newcommand{\uibkt}{Institut f\"ur Theoretische Physik, Universit\"at Innsbruck, Technikerstrasse 25, 6020 Innsbruck, Austria}
\newcommand{\iqoqi}{Institut f\"ur Quantenoptik und Quanteninformation, \"Osterreichische Akademie der Wissenschaften,Technikerstrasse 21A, 6020 Innsbruck, Austria\\ $^*$ These authors contributed equally to this work.\\
  $\dagger$ Present address: Fakult\"at f\"ur Physik, Ludwig-Maximilians-Universit\"at M\"unchen \& Max-Planck Institute
  of Quantum Optics, Germany}
\newcommand{\madrid}{Departamento de F\'isica Te\'orica I, Universidad Complutense, Avenida Complutense s/n, 28040 Madrid, Spain}
\hypersetup{colorlinks=false}
\begin{document}

\title{Quantum simulation of open-system dynamical maps with trapped
  ions}
  
\author{P. Schindler$^*$}
\affiliation{\uibke}
\author{M. M\"uller$^*$}
\affiliation{\madrid}
\author{D. Nigg}
\affiliation{\uibke}
\author{J. T. Barreiro$^{\dagger}$}
\affiliation{\uibke}
\author{E. A. Martinez}
\affiliation{\uibke}
\author{M. Hennrich}
\affiliation{\uibke}
\author{T. Monz}
\affiliation{\uibke}
\author{S. Diehl}
\affiliation{\uibkt}
\affiliation{\iqoqi}
\author{P. Zoller}
\affiliation{\uibkt}
\affiliation{\iqoqi}
\author{R. Blatt}
\affiliation{\uibke}
\affiliation{\iqoqi}

\begin{abstract}
Dynamical maps describe general transformations of the state of a physical system, and their iteration can be interpreted as generating a discrete time evolution. Prime examples include classical nonlinear systems undergoing transitions to chaos. Quantum mechanical counterparts show intriguing phenomena such as dynamical localization on the single particle level. Here we
  extend the concept of dynamical maps to an open-system,
  many-particle context: We experimentally explore the stroboscopic
  dynamics of a complex many-body spin model by means of a universal
  quantum simulator using up to five ions. In particular, we generate
  long-range phase coherence of spin by an iteration of purely
  dissipative quantum maps. We also demonstrate the characteristics of competition between 
  combined coherent and dissipative non-equilibrium evolution. This
  opens the door for studying many-particle non-equilibrium physics
  and associated dynamical phase transitions with no immediate
  counterpart in equilibrium condensed matter systems.
  An error detection and reduction toolbox that
  facilitates the faithful quantum simulation of larger systems is developed as a first step in this direction.
\end{abstract}

\maketitle

\tocless\section{Introduction}
Obtaining full control of the dynamics of many-particle quantum systems represents a fundamental scientific and technological challenge. Impressive experimental progress on various physical platforms has been made~\cite{maze08,ladd-nature-464-45,wrachtrup-jpcm-18-S807,clarke-nature-453-1031,obrien-science-318-1567,hanson-rmp-79-1217,schneider-repprogphys-75-024401,saffman-rmp-82-2313,bloch12,blatt12,guzik12,houck12}, complemented with the development of a detailed quantum control theory~\cite{bacon-pra-64-062302,lloyd-pra-65-010101,Lidar98,Baggio12}. Controlling the \textit{coherent} dynamics of systems well-isolated from the environment enables, for example, quantum computation in the circuit model~\cite{nielsen-book}. But this also allows for digital coherent quantum simulation with time evolution realized by sequences of small Trotter steps~\cite{lloyd-science-273-1073}, as demonstrated in recent experiments~\cite{lanyon_science-334-57,zhang12}. On the other hand, engineering the coupling of a system to its environment -- and thus its resulting \textit{dissipative} dynamics -- introduces new scenarios of dissipative quantum state preparation~\cite{PhysRevLett.77.4728,PhysRevLett.106.020504,PhysRevLett.106.090502,PhysRevLett.107.080503,barreiro-nature-470-486}, dissipative variants of quantum computing and memories~\cite{verstraete-nphys-5-633,PhysRevA.83.012304,kliesch-prl-107-120501} and non-equilibrium many-body physics~\cite{diehl-natphys-4-878,weimer-nphys-6-382,diehl-natphys-7-971}. Finally, complete control of both coherent and dissipative dynamics would enable the operation of an open-system quantum simulator, whose elementary building blocks have been demonstrated recently~\cite{barreiro-nature-470-486}, and which holds the promise to experimentally explore the dynamics of novel classes of non-equilibrium multi-particle quantum systems. 

The dynamics of these systems is often considered as continuous in time, described by many-body Lindblad master equations, cf. e.g.~\cite{zollerbook}. This may be conceived as a special instance of a more general setting, where a discrete time evolution of a system's reduced density matrix is generated by concatenated dynamical maps. So far, the concept of dynamical maps has proven useful for the description of periodically driven classical nonlinear systems~\cite{reichl-book}, and their quantum mechanical counterparts, such as the kicked rotor, providing one of the paradigmatic models of quantum chaos~\cite{chirikov79,izrailev90,haake-book}. Remarkable experiments have been performed with periodically driven systems of cold atoms, which have demonstrated some of the basic phenomena of quantum chaos such as dynamical localization~\cite{PhysRevLett.75.4598,PhysRevLett.80.4111,PhysRevLett.87.074102,
eurphyslett06}. At present all these studies are on the level of single particle physics.

\begin{figure*}[t]
\includegraphics[width=0.8\textwidth]{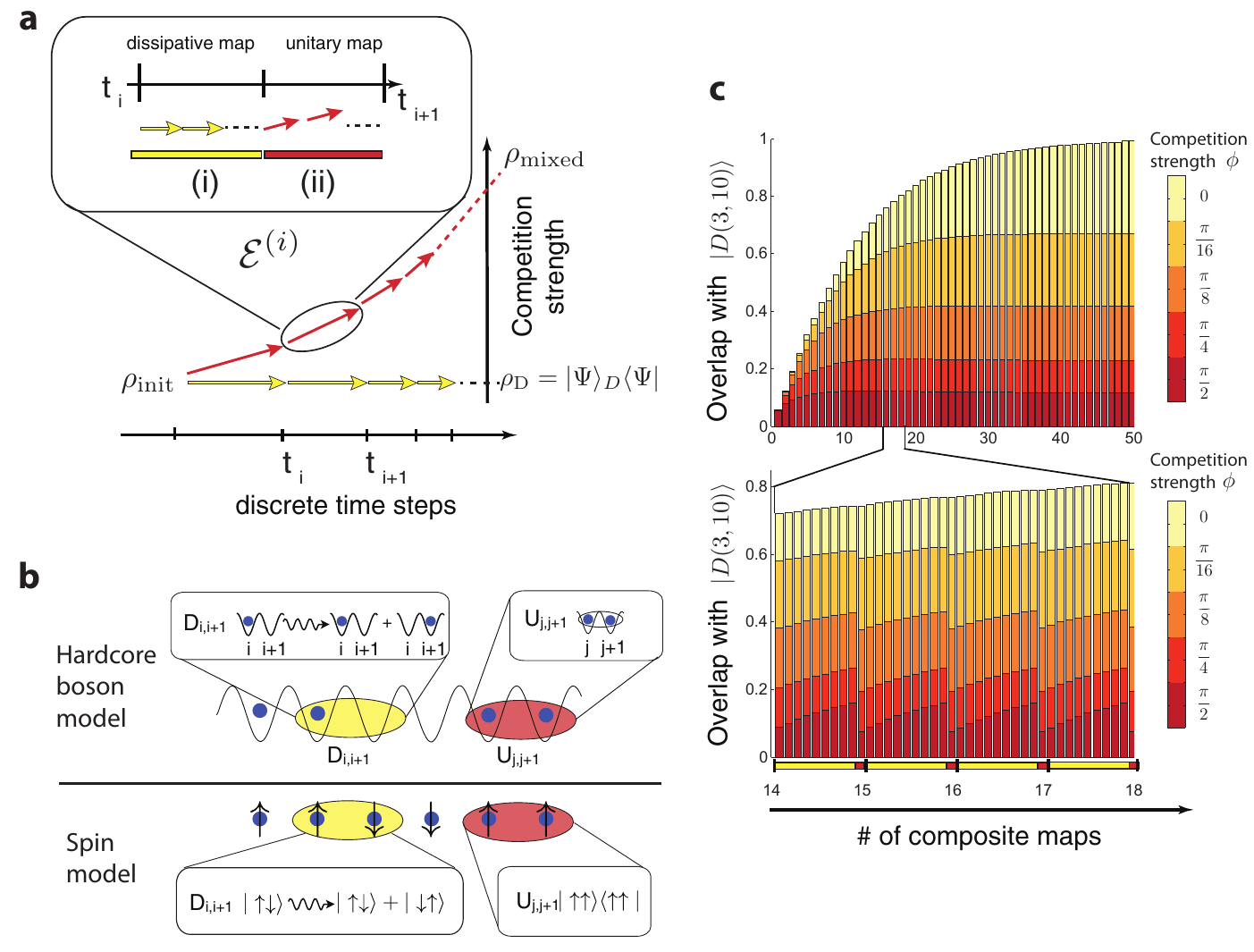}
\caption{\textbf{Competing dissipative and Hamiltonian dynamical maps in the spin or hardcore boson model.} \textbf{a,} Dynamical maps, acting on the reduced density matrix $\rho$ of an open many-body quantum system, can be composed of elementary (i) dissipative and (ii) unitary maps. The dissipative maps considered here drive the system into a pure, long-range ordered many-body ``dark'' state $\rho_D = |\psi\rangle_D \langle\psi |$. The addition of suitable Hamiltonian maps leads to a \emph{competition} of dissipative and coherent dynamics, in such a way that for strong enough coherent interactions, a phase transition into a mixed, disordered state results in large systems. \textbf{b,} Analogy of spins and hardcore bosons, in turn equivalent to bosons in the limit of low lattice filling (cf. Appendix~\ref{sec:bosonspin}): In the considered open-system spin-1/2 (hardcore boson) lattice model, spin excitations $|\!\uparrow \rangle$ ($|\! \downarrow \rangle$) can be identified with occupied (empty) lattices sites. Quasi-local dissipative maps $D_{i,i+1}$ acting on neighboring pairs of spins lead to delocalization of spin excitations (hardcore bosons) over pairs
 of lattice sites. Competing coherent dynamics is realized by unitary
 maps corresponding to interactions of  spin excitations (hardcore bosons) located on
 neighboring lattice sites $U_{j,j+1}$. \textbf{c,} Dynamics for an open-boundary chain of $N=10$ spins, starting in an initial state containing $m=3$ spin excitations (hardcore bosons) \si{\ref{sec:engineering}}. Time evolution is realized by
 sequentially applying composite maps consisting of (i) dissipative and (ii) coherent
 elementary dynamical maps, as shown in a. For vanishing Hamiltonian interactions, perfect long-range order (as measured by the
 overlap fidelity with the Dicke state $\ket{D(m,N)}$) is built up by
 the sequential application of dissipative maps (white), whereas increasing competing interactions (from yellow to red) lead to a decrease in the dissipatively created long-range order. The lower part of the figure shows a zoom into the composite maps 14 to 18, resolving that each of them is built up from 9 two-spin elementary dissipative maps, followed by 9 two-spin coherent maps, the latter which can be realized by a single, global unitary map in our setting.}
\label{fig:figure1}
\end{figure*}

In Sections II \& III below, we present a first experimental study of \emph{open-system many-particle} quantum dynamical maps for a complex spin model implemented in a linear ion trap quantum computing architecture using up to five ions. Building on recent advances in our experimental techniques, we harness the building blocks of our digital open-system quantum simulator device to concatenate elementary quantum maps with high fidelity. The controlled open-system maps are enabled by a digital simulation strategy, contrasting analog Hamiltonian quantum simulation with trapped ions~\cite{porras-prl-92-207901,friedenauer-nphys-4-757,kim-nature-465-590,islam-natcomm-2-377,bollinger12}. In this framework, we observe the key physical phenomena of multi-particle dissipative and coherent interactions in the stroboscopic system dynamics. In particular, we demonstrate the purely dissipative creation of quantum mechanical long-range phase order. Furthermore, we implement a competition between coherent and dissipative many-particle dynamics by alternating sequences of unitary and non-unitary maps outlined in Fig.~\ref{fig:figure1}a,
 and observe the destruction of phase coherence as a result. This reflects the hallmark feature of a strong coupling non-equilibrium phase transition predicted in a closely related driven-dissipative model of bosons~\cite{diehl10a}. 

The presented material is complemented by an extensive and self-contained collection of appendices, which provides further mathematical concepts and technical details.

\tocless\section{Competing dissipative and unitary dynamics in a complex spin model}
\label{sec:main_competing}
Two competing, non-commuting contributions to a Hamiltonian give rise to a quantum phase transition, if the respective ground states of each contribution separately favor states with different symmetries~\cite{sachdevbook}. The transition takes place at a critical value of the dimensionless ratio $g$ of the two competing energy scales. A non-equilibrium analog can be achieved in open many-body quantum systems, where coherent Hamiltonian and dissipative dynamics compete with each other: The role of the ground state is played by the stationary state of the combined evolution, and the dimensionless ratio $g$ is provided by a Hamiltonian energy scale vs. a dissipative rate. Such a situation has been addressed previously theoretically in the context of driven-dissipative dynamics of atomic bosons on a lattice~\cite{diehl10a}: A dissipative dynamics can be devised to drive the system from an arbitrary initial state with density matrix $\rho_{\text{in}}$ into a Bose-Einstein condensate with long-range phase coherence as the unique, pure ``dark'' state $|\psi\rangle_D$ of the dissipative evolution, i.e. $\rho_{\text{in}} \to\rho_D= |\psi\rangle_D \langle\psi |$ for long enough waiting time. Supplementing this dynamics with a Hamiltonian representing local interactions, being incompatible with the dissipative tendency to delocalize the bosons, gives rise to a strong coupling dynamical phase transition. It shares features of  a quantum phase transition in that it is driven by the competition of non-commuting quantum mechanical operators, and a classical one in that the ordered phase terminates in a strongly mixed state.

In our experiment, we consider analogous open-system dynamics of a quantum spin-1/2 -- or hardcore boson -- model, realized with trapped ions. A schematic overview of the relation of the ionic spin- and the atomic boson model is given in Fig.~\ref{fig:figure1}b, whereas a more detailed description is provided in Appendices~\ref{sec:bosonspin} and~\ref{sec:engineering}. The discrete time evolution is generated by sequences of dynamical or Kraus maps $ \mathcal{E}^{(l)}$ acting on the system's reduced density matrix $\rho$ as illustrated in Fig.~\ref{fig:figure1}a, with time steps $t_\ell \to t_{\ell + 1}$ represented by 
\begin{eqnarray*}\label{eq:dynmap}
\rho  (t_\ell ) \mapsto \rho (t_{\ell+1})  = \mathcal{E}^{(l)}[\rho(t_\ell)] = \sum_{k=1}^K E_k^{(l)} \rho(t_\ell)  E_k^{(l)\dag}.
\end{eqnarray*}
The set of Kraus operators $\{E_k^{(l)} \}$ satisfies $\sum_{k=1}^K E_k^{(l)\dag} E_k^{(l)} =1$~\cite{nielsen-book}. While the familiar sequences of unitary maps are obtained for a single Kraus operator $K=1$, dissipative dynamics corresponds to multiple Kraus operators $K>1$. In particular, the continuous time evolution of a Lindblad master equation is recovered in the limit of infinitesimal time steps, cf. Appendix~\ref{sec:mastereq}.  The dissipative dynamics studied in our spin model 
is governed by dynamical maps according to two-body
Kraus operators acting on pairs of neighboring spins $i,i+1$:
\begin{eqnarray}
E_{i,1} = c_i,\quad\quad
E_{i,2} = 1 - c_i^\dag c_i.\label{eq:kraus}
\end{eqnarray}
The elementary operators generating the dynamics are given by
\begin{eqnarray}\label{eq:jump}
c_i = (\sigma_i^+ + \sigma_{i+1}^+ )  (\sigma_i^- - \sigma_{i+1}^- ),
\end{eqnarray}
where $\sigma_i^{\pm} = (\sigma_i^x \pm i \sigma_i^y)/2$ are spin-1/2 raising and lowering operators acting on spin $i$. In the continuous time limit, the operators $c_i$ correspond precisely to Lindblad quantum jump operators and generate a dissipative evolution described by a quantum master equation, cf. Appendix~\ref{sec:mastereq}. The operators act bi-locally on pairs of spins, as visualized in Fig.~\ref{fig:figure1}b: Physically, they map any antisymmetric component in the wave function on a pair of sites into the symmetric one, or -- in the language of hardcore bosons -- symmetrically delocalize particles over pairs of neighboring sites. Since this process takes place on each pair of neighboring sites, eventually only the symmetric superposition of spin excitations over the whole array persists as the stationary state of the evolution: Iteration of the dissipative dynamical map attracts the system towards a unique dynamical fixed point, or dark state, characterized by $\rho(t_{\ell+1})  = \rho(t_\ell) \equiv \rho_D$, resulting from the property $c_i \rho_D =0$ for all $i$ separately. More specifically, for $m$ spin excitations initially present in the array of $N$ spins, this pure dark state is given by the Dicke state
\begin{eqnarray*}
\ket{\psi}_D = \ket{D(m,N)} \sim 
\Big( \sum_{i=1}^N \sigma_i^+ \Big)^m \ket{\!\qzero}^{\otimes N}
\end{eqnarray*}
with $m$ \emph{collective} spin excitations. The delocalization of the spin excitations over the whole array gives rise to the creation of entanglement and quantum mechanical off-diagonal long range order, witnessed, e.g., by the single-particle correlations $\langle \sigma^+_i \sigma^-_j\rangle \neq 0$ for $|i-j|\to \infty$ \si{\ref{sec:experimental}}.

In our interacting lattice spin system, competing unitary dynamics can be achieved by the stroboscopic realization of coherent maps $\rho  (t_\ell ) \mapsto \rho (t_{\ell+1}) = U  \rho  (t_\ell ) U^\dag$ with $U =\exp (-\mathrm i \phi H)$ according to the dimensionless spin Hamiltonian 
\begin{eqnarray*}
H = \sum_i \ket{\!\uparrow}\bra{\uparrow\!}_i \otimes  \ket{\!\uparrow}\bra{\uparrow\!}_{i+1} = \sum_i (1+ \sigma^z_i) (1+ \sigma_{i+1}^z)/4.
\end{eqnarray*} 
The bi-local terms of the Hamiltonian describe interactions of spin
excitations or hardcore bosons located on neighboring sites (see
Fig.~\ref{fig:figure1}a). The competition between dissipative and
unitary dynamics arises since the dissipative dark states $\rho_D$ are
not eigenstates of the Hamiltonian, which is diagonal in Fock space
and thus leads to a dephasing of the dissipatively induced
off-diagonal order. The value of the angle $\phi \in [0 , \pi/2]$
determines the strength of the competition between the Hamiltonian and
dissipative dynamics and plays a role analogous to a dimensionless
ratio of energy scales in a quantum phase transition, or of an energy
scale and a dissipative rate in the above scenario. Clearly, for small
system sizes, the sharp transition found in the thermodynamic limit is
replaced by a smooth crossover as indicated in
Fig.~\ref{fig:figure1}c.

\tocless\section{Experimental realization}
\begin{figure}[tbp]
\includegraphics[width=0.46\textwidth]{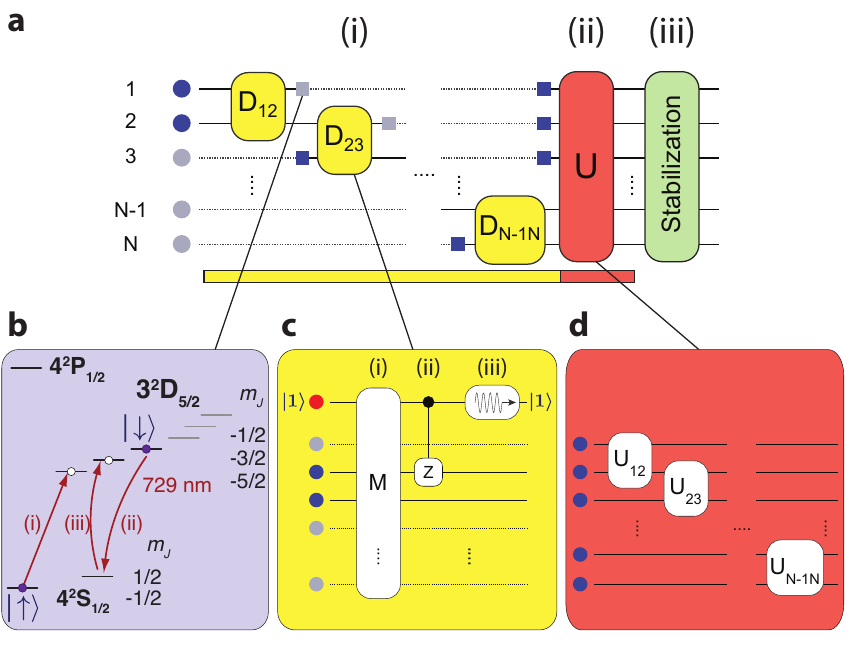}
\caption{\textbf{Experimental procedure to implement open-system
    dynamical maps.} \textbf{a}, Schematic overview of the
  experimental implementation of a composite dynamical map consisting
  of (i) multiple elementary dissipative maps, (ii) coherent
  competition, and (iii) error detection and correction. Decoupled
  ions are represented as gray bullets and decoupling (re-coupling)
  operations as gray (blue) squares. \textbf{b}, Scheme for decoupling
  ions from the interaction with the manipulating light fields: (i)
  shelve population from $4S_{1/2}(m=-1/2)=\ket{\!\!\qone}$ to
  $3D_{5/2}(m=-5/2)$, (ii) transfer the population from
  $3D_{5/2}(m=-1/2)=\ket{\!\!\qzero}$ to $4S_{1/2}(m=+1/2)$, and
  subsequently to (iii) $3D_{5/2}(m=-3/2)$. \textbf{c}, A single
  dissipative element is realized using two system spins and one
  ancilla qubit
  ($\ket{\azero}\equiv\ket{\!\!\qzero}$,$\ket{\aone}\equiv\ket{\!\!\qone}$)
  by (i) mapping the information whether the system is in the
  symmetric or antisymmetric subspace onto the logical states
  $\ket{\aone}$ or $\ket{\azero}$ of the ancilla, respectively; (ii)
  mapping the antisymmetric onto the symmetric state using a
  controlled phase flip conditioned by the state of the ancilla qubit;
  and finally (iii) reinitialization of the ancilla qubit via optical
  pumping using the $4^2P_{1/2}$ state \si{\ref{sec:SI_reinitialization}}. \textbf{d}, Schematic view of the competing interaction
  consisting of quasi-local unitary maps $U_{j,j+1}$.}
\label{fig:figure3}
\end{figure}

The simulation is performed with
$^{40}$Ca$^+$ ions, confined to a string by a macroscopic linear Paul
trap \si{\ref{sec:experimental}}.  Each ion hosts a qubit or spin-1/2, which is encoded in the
$4S_{1/2} (m=-1/2) \; = \ket{1}\equiv\ket{\!\!\qone}$ and the $3D_{5/2} (m
=-1/2) \; = \ket{0}\equiv\ket{\!\!\qzero}$ states. The backbone of this
digital quantum simulator setup is a universal set of high-fidelity
operations, which are realized by exactly timed laser pulses resonant
with the qubit transition \si{\ref{sec:experimental}}.  The entangling gate operations
\cite{molmer-prl-82-1835} act on the entire ion string, but the
elementary dissipative maps $D_{i,i+1}$ act on only two of the $N$
system spins (Fig.~\ref{fig:figure3}a). We achieve operations on
subsets of ions via decoupling all ions \textit{not} involved in the
elementary map, by shelving their population into additional storage
states (Fig.~\ref{fig:figure3}b). In these electronic states, these
ions are effectively ``inactive'' as they do not interact with the
globally applied laser beams. This novel spectroscopic decoupling
technique is experimentally simpler than physically moving the ions
with respect to the laser beam~\cite{wineland_scaling}.

To observe the complex dynamics of the open interacting spin system,
we combine these experimental techniques (i) to generate long-range
phase coherence of spins by an iteration of purely dissipative maps,
(ii) to combine these dissipative dynamics with competing coherent
maps, (iii) and finally to implement quantum non-demolition (QND)
readout and quantum-feedback protocols for error detection and
stabilization.

(i) The basis of the composite maps is a single elementary dissipative
map $D_{i,i+1}$ that is implemented by a quantum circuit of coherent
gate operations and addressed optical pumping (see
Fig.~\ref{fig:figure3}c), acting on the two currently active ions
$i,i+1$ and an ancillary qubit, which is used to engineer the coupling
to the environment
\cite{barreiro-nature-470-486,lloyd-pra-65-010101,bacon-pra-64-062302}. The
circuit decomposition of the three-qubit unitary underlying a single
elementary dissipative map, is obtained from an optimal control
algorithm \si{\ref{sec:numerical_optimization}}, resulting in a sequence of 17 operations
containing 4 entangling gates. We have characterized a single
elementary dissipative map by quantum process tomography on the two
system qubits leading to a mean state fidelity of 68(1)\%, 
which
approximately corresponds to an average fidelity of over 98\% per gate
operation \si{\ref{sec:experimental}}. Due to the considerable complexity of the gate
sequence, errors occur in different bases and are expected to average
out and give rise to depolarizing noise without any preferred
direction. Therefore, the actual implemented dynamics can be modeled
as a combination of the ideal, tailored dissipative map and the
depolarizing noise channel. Detailed numerical simulations show that
the error is mainly caused by laser frequency and magnetic field
fluctuations \si{\ref{sec:implementationmap}}.

\begin{figure}[tbp]
\includegraphics[width=0.46\textwidth]{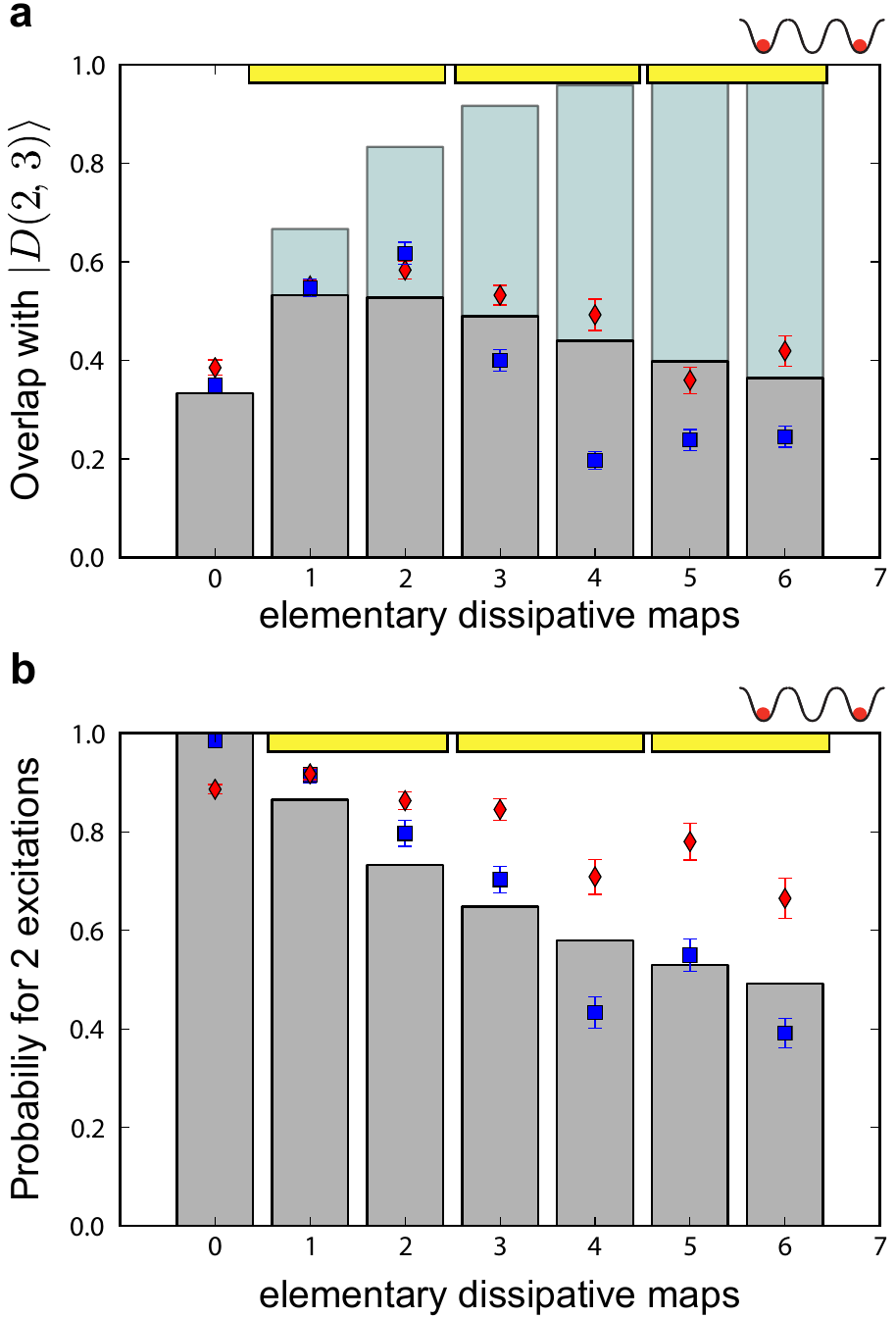}
\caption{\textbf{Experimental results of dissipatively induced
    delocalization through composite dynamical maps with 3+1 ions.}
  The results from an ideal model are shown in light-blue bars whereas
  those from a model including depolarization noise are indicated by
  dark-grey bars. Blue rectangles indicate the experimentally observed
  dynamics without any correction scheme whereas red diamonds
  include a post-selective error detection scheme (error bars,
  1$\sigma$).  {\textbf a}, Dissipative pumping into a three-spin
  Dicke state: Starting in an initial product state with two localized
  spin excitations $\ket{\qone\qzero\qone}$, the application of the first two
  elementary dissipative maps leads to an increase in the
  delocalization of the two excitations over the spin chain, which is
  reflected by an increasing state overlap fidelity with the
  three-spin Dicke state $\ket{D(2,3)}$. However, after applying a
  second and a third composite dissipative map, a decrease in the
  state overlap fidelity sets in and becomes dominant for long
  sequences of dynamical maps. {\textbf b}, The presence of
  depolarizing noise results in population leakage out of the initial
  subspace with $m=2$ spin excitations. This effect is evident in the
  decay of the probability of finding the three-spin system in the
  $m=2$ excitations subspace as a function of the number of applied
  elementary dissipative maps. A single composite  dissipative map is
  indicated by a yellow bar. 
 }\label{fig:figure4} \end{figure}

We repeatedly apply such elementary maps $D_{i,i+1}$, interspersed with
decoupling pulses to coherently transfer ions in and out from the
storage states, to build up the composite dissipative map in a modular
way, Fig.~\ref{fig:figure3}a. Using 3+1 ions, we studied pumping
towards Dicke states in a three-spin chain with open boundary
conditions, where we applied up to three simulation timesteps, each
consisting of the two elementary maps $D_{1,2}$ and $D_{2,3}$. The
results displayed in Fig.~\ref{fig:figure4}a show a clear experimental
signature of dissipatively induced delocalization of the spin
excitations during the application of the first two elementary
dissipative maps, before experimental imperfections become dominant
for longer sequences.

\begin{figure}[tbp]
\includegraphics[width=0.46\textwidth]{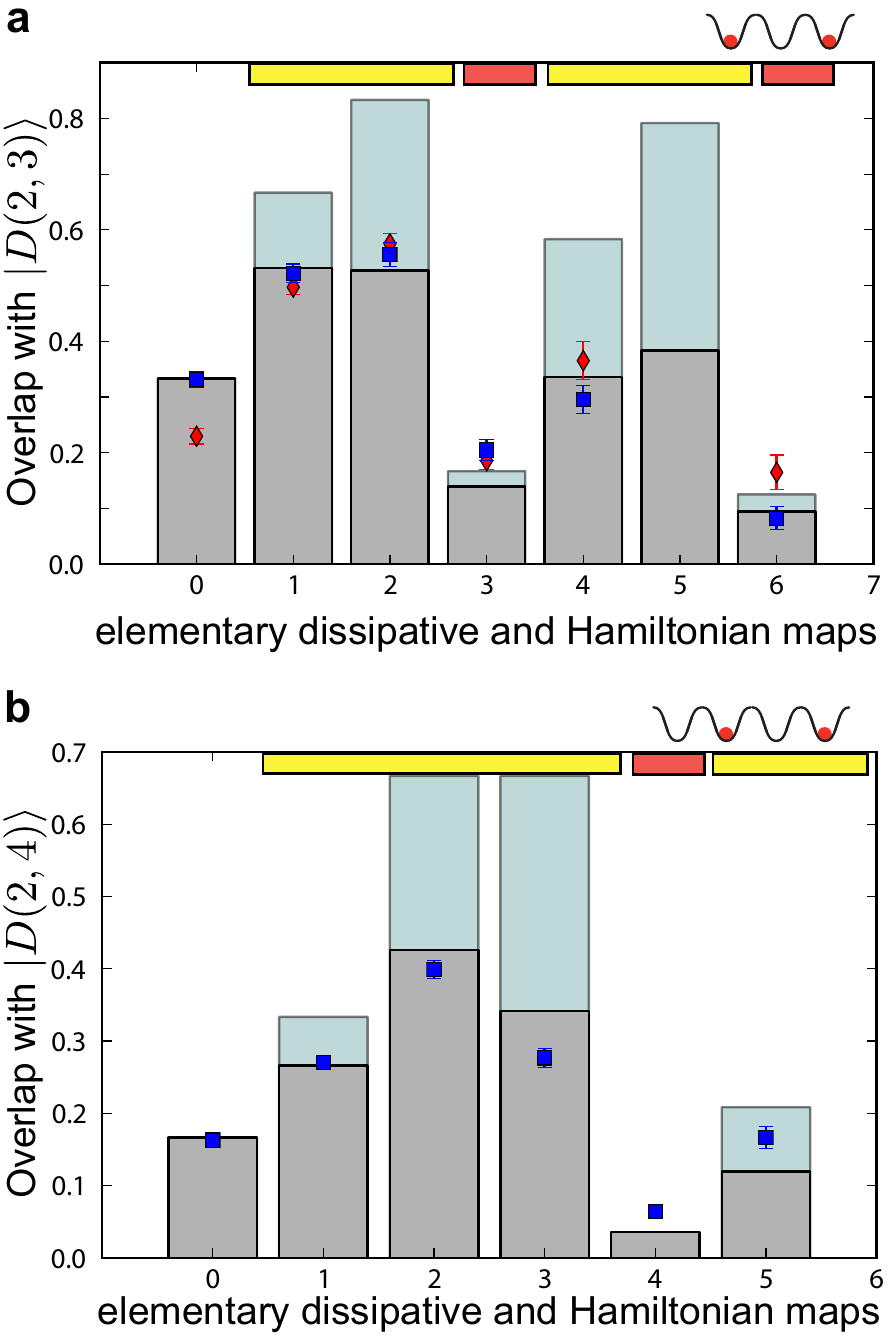}
\caption{\textbf{Experimental results for competing dissipative and
    coherent dynamics with 3+1 and 4+1 ions.} As in
  Fig.~\ref{fig:figure4}, the results from an ideal model are shown in
  light-blue bars whereas those from a model including depolarization
  noise are indicated by dark-grey bars. Blue rectangles indicate the
  experimentally observed dynamics without any correction scheme
  whereas red diamonds include a post-selective error detection
  scheme (error bars, 1$\sigma$). The application of dissipative
  (coherent) maps is indicated by yellow (red) bars.  Competing
  dissipative and coherent dynamics for $m=2$ excitations in chains of
  {\textbf{a,}} $N=3$ and {\textbf{b,}} $N=4$ spins: the spin chains
  are first driven towards the Dicke-type dark state by the two and
  three elementary dissipative maps for a system of 3 and 4 spins. The
  subsequent application of the non-compatible unitary dynamical maps
  leads to a strong decrease of the overlap with the respective Dicke
  states, before subsequent elementary dissipative maps again start to
  pump the system back towards the Dicke states.
}  \label{fig:figure42} \end{figure}

(ii) To investigate the competition between dissipative and competing
Hamiltonian dynamics, we added elementary unitary maps according to
nearest-neighbor spin-spin interactions to the composite dynamical
maps \si{\ref{sec:engineering}}. Note that due to the commutativity of the two-spin
elementary unitary maps, the composite, globally acting unitary map
can be realized by a single unitary sequence
(Fig.~\ref{fig:figure3}d). The results displayed in
Fig.~\ref{fig:figure42} for experiments with 3+1 and 4+1 ions show a
clear fingerprint of incompatible Hamiltonian dynamics, which competes
with the dissipative maps driving the spin chains towards the Dicke
states. Further measurements with varying excitation number and
competition strength are discussed \si{\ref{sec:additionalexp}}.

\begin{figure*}[tbp]
\includegraphics[width=0.90\textwidth]{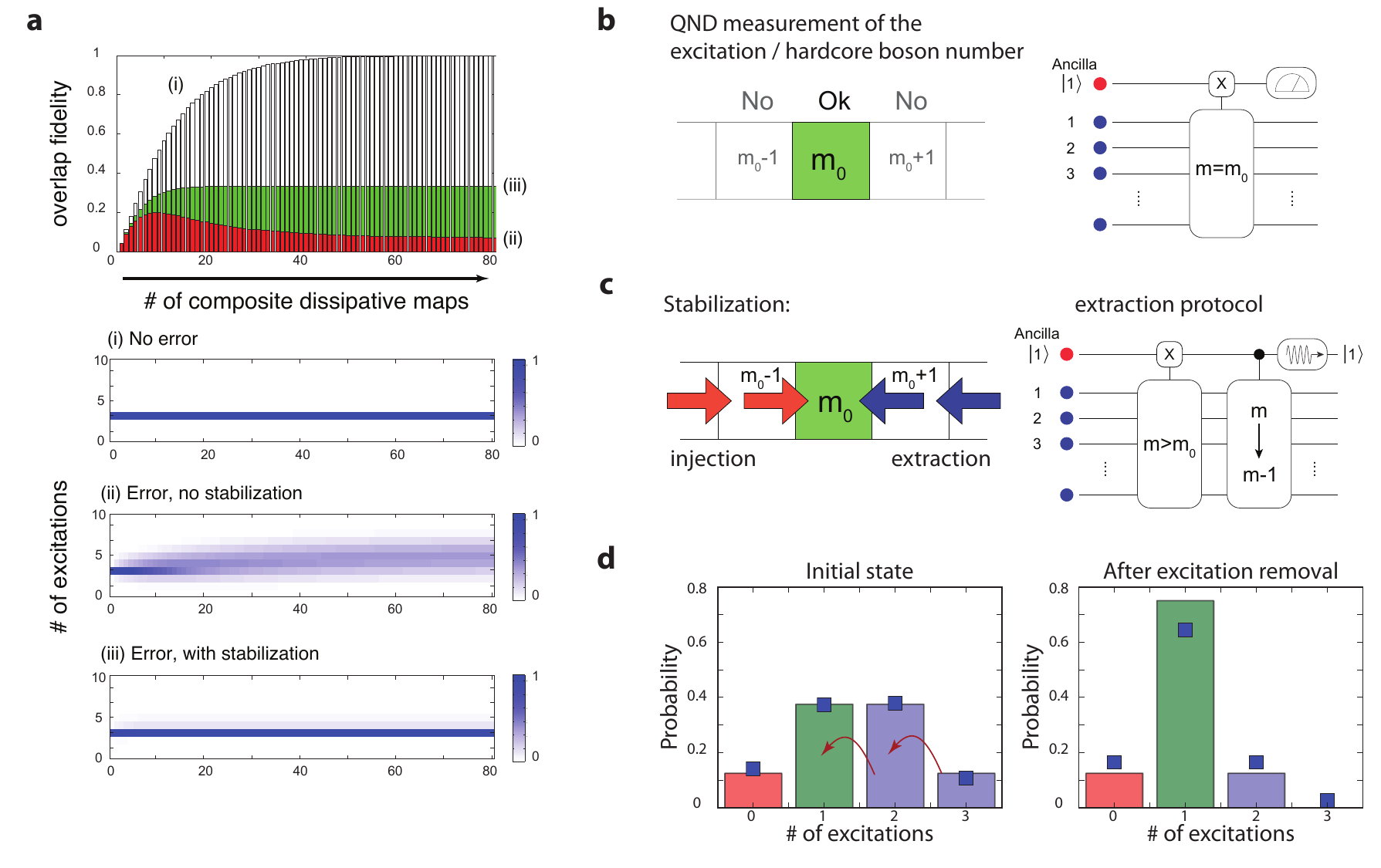}
\caption{{\bf{Experimental error detection and reduction techniques.}}
  {\textbf{a,}} Numerical simulation of the time evolution consisting
  of composite dynamical maps building up long-range order quantified
  by overlap fidelity with the Dicke state $\ket{D(m,N)}$, starting in
  an initial state containing $m=3$ spin excitations on an
  open-boundary chain of $N=10$ spins. The simulation was performed
  for (i) ideal operations without experimental imperfections, (ii) a
  model including experimental errors, (iii) a model including errors
  but with the stabilization scheme.  The lower part depicts the
  expected excitation distribution of the time evolution demonstrating
  that the stabilization keeps the system in the correct excitation
  subspace.  {\textbf b,} Schematic idea and quantum circuit for a
  post-selective QND measurement of the total spin excitation number:
  First the information whether the $N$ system spins are (not) in the
  subspace with $m$ spin excitations or not is mapped coherently onto
  the logical state $\ket{\aone}$ ($\ket{\azero}$) of an ancillary
  qubit. A subsequent projective measurement of the latter indicates
  the presence of an erroneous excitation number, in which case the
  experimental run is discarded.  {\textbf c,} Active QND feedback
  scheme to stabilize the spin system in a desired excitation number
  subspace by actively extracting (injecting) spin excitations in case
  errors in previous dynamical maps have led to a larger (smaller)
  excitation number than present initially. The information whether
  more excitations than the desired value are present in the system
  (or not) is coherently mapped onto the state $\ket{\aone}$
  ($\ket{\azero}$) of the ancilla qubit. A non-unitary
  controlled-operation, only active for the ancilla qubit in
  $\ket{\aone}$, then extracts in a minimally destructive way one spin
  excitation from the system and automatically stops once this is
  achieved. {\textbf{d,}} Experimental demonstration of the
  stabilization protocol for $m_0=1$ using 3+1 ions by applying the
  excitation removal to an initial state consisting of an equal
  superposition of all computational states, $|\psi_0\rangle =
  1/\sqrt{8} (|\qzero \rangle + |\qone \rangle)^{\otimes 3}$. Here, a
  single spin excitation should be removed if two or more excitations
  are present in the system (blue bars). Thus, pumping from the $m=3$
  into the $m=2$ and from the $m=2$ into the $m=1$ subspace is
  expected, whereas population initially present in the $m=0$ and
  $m=1$ subspaces should be left untouched. The correct populations
  after the application of the excitation removal protocol are
  confirmed by the measured data (blue rectangles) which are close to
  the ideally expected behavior (colored bars). Data error bars are
  smaller than the markers. } \label{fig:figure5} \end{figure*}

(iii) Under the ideal dynamical maps, the total spin excitation (or
hardcore boson) number $m$ is a conserved quantity, however the
presence of depolarizing noise as the dominant error source results in
population leakage out of the initial excitation number subspace. To
reduce this detrimental effect and enable the implementation of longer
sequences of dynamical maps, we developed and implemented two
counter-strategies (see Fig.~\ref{fig:figure5} for details). In a
first approach we applied a quantum non-demolition (QND) measurement
of the spin excitation number at the end of the sequence of dynamical
maps, which allowed us to detect and discard experimental runs with a
final erroneous excitation number and thereby improve the overall
simulation accuracy. This global measurement is QND in the sense that
only information about the total number of excitations, but not on
their individual spatial locations along the chain is acquired; thus
the simulation subspace is not disturbed \si{\ref{sec:QND_postselection}}. The results shown in
Fig.~\ref{fig:figure4} and Fig.~\ref{fig:figure42} confirm that the
errors in the spin excitation number are strongly suppressed and a
reasonable overlap with the ideal evolution can be maintained for more
simulation time steps.

Complementary to this post-selective method, we introduced a more
powerful, active QND feedback scheme, which bears similarities to
quantum feedback protocols as realized with photons in a cavity
\cite{haroche_feedback}. The key idea is to actively stabilize the
spin system during the sequence of dynamical maps in a subspace with a
particular spin excitation (or hardcore boson) number \si{\ref{sec:error_reduction}}. In order
to be able to perform this stabilization with a single ancilla qubit,
we break the stabilization process into two independent parts, where
the first part removes one excitation if there are too many
excitations in the system, and the second part adds one excitation if
needed. Similarly to the post-selective technique presented above,
first the information whether there are too many (few) excitations in
the system is coherently mapped onto the ancilla
qubit. 
Depending on the state of the ancilla qubit, a single excitation is
removed from (injected into) the system by a quantum feedback
protocol. This extraction (injection) is in general an ambiguous
process, as the excitation can be removed (injected) on multiple
sites. We use a scheme that tries to perform the removal (injection)
subsequently on each site and stops once it was successful. Using only
a single ancilla qubit, this process cannot be performed efficiently
as a unitary process, therefore we developed a technique making use of
the resetting and decoupling techniques described above \si{\ref{sec:error_reduction}}.

We demonstrate the excitation removal for a chain of 3+1 spins,
initially prepared in an equal superposition of all basis states, as
shown in Fig.~\ref{fig:figure4}c. At the current level of experimental
accuracy, the implementation of this stabilization scheme cannot
improve the performance when used in the full simulation sequence \si{\ref{sec:exp_stabilization}}. We emphasize, however, that our approach relies only on a single
ancillary qubit, regardless of the system size. More generally, such
customized error detection and reduction strategies will incur a
substantially reduced resource overhead as compared with full-fledged
quantum error correction protocols. For the future, we therefore
envision that methods as those demonstrated here will be a promising
practical avenue to considerably extend the runtime of quantum
simulators.

\renewcommand{\refname}{}
%

\begin{center}\textbf{Acknowledgments}\end{center}

We gratefully acknowledge support by the Austrian Science Fund (FWF),
 through the SFB FoQus (FWF Project No. F4002-N16 and F4016-N16) and
 the START grant Y 581-N16 (S. D.), by the European Commission
 (AQUTE), by IARPA as well as the Institut
 f\"{u}r Quantenoptik und Quanteninformation GmbH.  M. M. acknowledges
 support by the CAM research consortium QUITEMAD S2009-ESP-1594,
 European Commission PICC: FP7 2007-2013, Grant No. 249958, and the
 Spanish MICINN grant FIS2009-10061.

\begin{center}\textbf{AUTHOR CONTRIBUTIONS}\end{center}

M.M., P.S., J.T.B. and S.D. developed the research, based on
theoretical ideas conceived with P.Z.; P.S. and D.N. performed the
experiments; P.S. and T.M. analysed the data; P.S., J.T.B., D.N.,
T.M., E.A.M., M.H. and R.B. contributed to the experimental set-up;
P.S., M.M. and S.D wrote the manuscript, with revisions provided by
J.T.B., P.Z. and R.B..; all authors contributed to the discussion of
the results and manuscript.

$$\phantom{\int}\\\phantom{\int}$$

$$\phantom{\int}\\\phantom{\int}$$

$$\phantom{\int}\\\phantom{\int}$$

$$\phantom{\int}\\\phantom{\int}$$

\clearpage
\appendix

\begin{center}\hspace{-4mm}\textbf{APPENDICES}\end{center}
\begin{minipage}[l]{0.95\columnwidth}
	\begingroup
	\tableofcontents
	\endgroup
\end{minipage}
\vspace{5mm}
\renewcommand{\thefigure}{\textbf{A\arabic{figure}}}
\setcounter{figure}{0}


These appendices provide background material on details of the theory
and the experimental implementation of the quantum simulation of open
system dynamical maps with trapped ions discussed in the main text. In
addition, it covers further experimental simulation results which are not discussed in the main
text. Appendix~\ref{sec:mastereq} discusses how the continuous time
evolution of a Lindblad master equation is recovered in the limit of
infinitesimal time steps of Kraus maps. Appendix~\ref{sec:bosonspin}
shows how the dissipative spin operators are constructed in analogy to
the case of bosons.  Appendix~\ref{sec:experimental} provides details
on the experimental system and the available set of tools for
open-system quantum simulation. In Appendix~\ref{sec:engineering} we
present in more detail the general concept of the engineering of
elementary dissipative and coherent Kraus maps, and also provide the
specific sequences of operations that have been used in the
experimental implementation. In
Appendix~\ref{sec:exp_characterization} we discuss the quantitative
experimental characterization of elementary dissipative Kraus
maps. Based on these measurements, we describe in more detail a
theoretical noise model, which captures the dominant experimental
imperfections, and present analytical expressions and information on
the numerical simulations for the dynamics of the open many-body spin
system under dynamical maps. Appendices~\ref{sec:QND_postselection}
and~\ref{sec:error_reduction} contain background information on the
error detection and reduction schemes, as well as experimental details
on their implementation. Finally, in Appendix~\ref{sec:additionalexp}
we discuss additional experimental simulation results, including the
realization of competing dissipative and coherent dynamical maps for
varying Hamiltonian interaction strengths and initial states, and the
combined implementation of dissipative dynamical maps with the error
reduction protocol.

Table~\ref{tab:Resources} provides an overview of the algorithmic building blocks required for the implementation of elementary dissipative Kraus maps, the realization of Hamiltonian dynamical maps, spectroscopic decoupling of ions, and elements for the detection and reduction of errors in our ion-trap quantum simulator. The table describes the experimental resources needed for the implementation of these elementary sub-routines as well as for the combination of these building blocks in more complex composite dynamical maps. The details of the underlying Kraus map engineering as well as the detailed quantum circuits used in the experiment are presented in the corresponding appendices below.
\begin{table*}
  \centering
  \begin{tabular}{|c|c|c|c|c|c|c|}
    \hline
     & Nr.~of & Nr.~of global & Nr.~of AC- & Nr.~of & Total number & Described  \\
     Algorithm type & MS gates & rotations $R$ & Stark shifts $S_Z$ & resets & of operations & in Appendix\\
    \hline
    Elementary dissipative map & 6 & 7 &9 &1& 23 & \ref{sec:supp_diss_Kraus_maps} \\
    Hamiltonian competing dynamics with 3+1 ions & 2 & 3 & 2 & 0 & 7 & \ref{sec:maps}\\
    Hamiltonian competing dynamics with 4+1 ions & 3 & 4 & 4 & 0 & 11 & \ref{sec:maps}\\
    QND post-selective error detection & 4 &8 & 6 & 0&18 & \ref{sec:QND_postselection} \\
    Mapping for the spin excitation removal & 2 & 9 & 4 & 0&15 & \ref{sec:error_reduction} \\
    Mapping for the spin excitation injection & 3 & 12 & 7 &0& 22 & \ref{sec:error_reduction} \\
    Excitation injection / removal step (single site) & 4 & 0 & 2 & 0 & 6 & \ref{sec:error_reduction}\\
    Spectroscopic decoupling & 0 & 5 & 4 &  0 & 9 & \ref{sec:decoupling}\\
    \hline
    Composite dissipative map (3 spins)& 12&26 &24 &2 &64 &\\
    Composite diss. and coh. dyn. map (3 spins) & 14&29 &26 &2 &71 &\\
    Composite diss. and coh. map + QND (3 spins) & 18&40 &27 &3 &88& \\
    \hline
    Composite dissipative map (4 spins) & 18&33 &35 &3 &89& \\
    Composite diss. and coh. dyn. map (4 spins) & 21&44 &44 &3 &112& \\

    \hline
  \end{tabular}
  \caption{Summary of the required resources for the elementary and composite dynamical maps and additional tools used 
    in the quantum simulation. The required operations for the composite maps do not strictly match
    the sum of the required elementary operations, since in the implementation of composite maps synergy effects in the resources for the spectroscopic decoupling operations can be exploited.}
  \label{tab:Resources}
\end{table*}


\section{Dynamical maps vs. quantum master equation}
\label{sec:mastereq}
The dissipative Kraus maps specified in equation (\ref{eq:kraus}) are
obtained as a special case of the operators $E_{i,1} =\sin \theta
c_i$, $E_{i,2} = 1 + (\cos \theta - 1) c_i^\dag c_i$, for $\theta =
\pi/2$. This limit corresponds to a deterministic action of the Kraus
map \si{ II}, in this case generating truly stroboscopic
dynamics. Instead, in the limit $\theta \to 0$, we approximate
$E_{i,1} \approx \theta c_i$, $E_{i,2} \approx 1-\frac 12 \theta^2
c_i^\dagger c_i$.
 In this limit, the sequence of dynamical maps reduces to the
 continuous time evolution described by a quantum master equation
 entirely generated by a dissipative Liouville operator, $\mathcal L[\rho] = \sum_i \left(  c_i \rho c_i^\dag - \frac{1}{2} \{ c_i^\dag c_i,\rho\} \right)$~\cite{nielsen-book}. Similarly, the Hamiltonian Kraus map can be expanded, $\exp
 (- \mathrm i \phi H) \approx 1 - \mathrm i \phi H$ for $\phi \to 0$. The dynamics in this continuum limit is then described by the 
 quantum master equation $\partial_t \rho = - \mathrm i [UH,\rho] +\kappa \mathcal L [\rho]$, with the dimensionful energy scale $U$ ($U \mathrm{d}t = \phi$) and dissipative rate $\kappa$ ($\kappa \mathrm{d}t =
 \theta^2$). Here, d$t$ is the physical time required for the implementation of one Kraus map in the digital simulation. The dimensionless ratio describing the competition $g = U/\kappa = \phi/\theta^2
 $ remains well-defined in this limit. We finally note that a
 temporal coarse graining implemented by averaging over a sequence of
 elementary maps with even larger discrete mapping steps gives rise to a
 quasi-continuous evolution, as shown numerically in
 Fig. \ref{fig:figure1}c. In a large system, the quasilocal operations
 can be coarse grained also over space. The result is an effective
 quasi-continuous master equation dynamics for the density operator.

\section{Atomic boson vs. ionic spin model}
\label{sec:bosonspin}
The dissipative spin operators $c_i$ of
 equation (\ref{eq:jump}) are constructed in complete analogy to the
 case of bosons, which has been proposed theoretically in~\cite{diehl-natphys-4-878}: Formally, and as further detailed in
 Appendix~\ref{sec:engineering}, they obtain by replacing the spin raising (lowering) operators
 $\sigma_i^+ (\sigma_i^-)$ by bosonic creation (annihilation)
 operators $a^\dag_i (a_i)$ of atoms confined to an optical
 lattice. In that case, the dark state is given by $m$ symmetrically
 delocalized particles on $N$ lattice sites, i.e. $\ket{\psi}_D =
 (m!)^{-1/2} \left(\sum_{i=1}^N a_i^\dag \right)^m\ket{0}^{\otimes N}$
 ---~the Dicke dark states of the spin model are replaced by a fixed
 number Bose-Einstein condensate (the bosonic vacuum state is defined
 by $a_i \ket{0}^{\otimes N}=0$ for all $i$). Using the
 Holstein-Primakoff representation of spin $1/2$ operators in terms of
 bosons, $\sigma_i^+ = a^\dag_i \sqrt{ 1 - \hat n_i} $ ($\hat n_i =
 a_i^\dag a_i$), it is seen that the dissipative spin operators reduce
 to their bosonic counterpart in the limit of small average occupation
 $\bar n = \langle \hat n_i \rangle \ll 1$, where the square root can
 be safely replaced by one.

\section{Experimental system and techniques}
\label{sec:experimental} 
In this appendix we will describe the available operations in our
universal ion-trap quantum simulator.

\subsection{Coherent gates}
\label{sec:SI_coherent_gates}
\begin{figure}[t!]
\center
\includegraphics[angle=0,width=0.8\columnwidth]{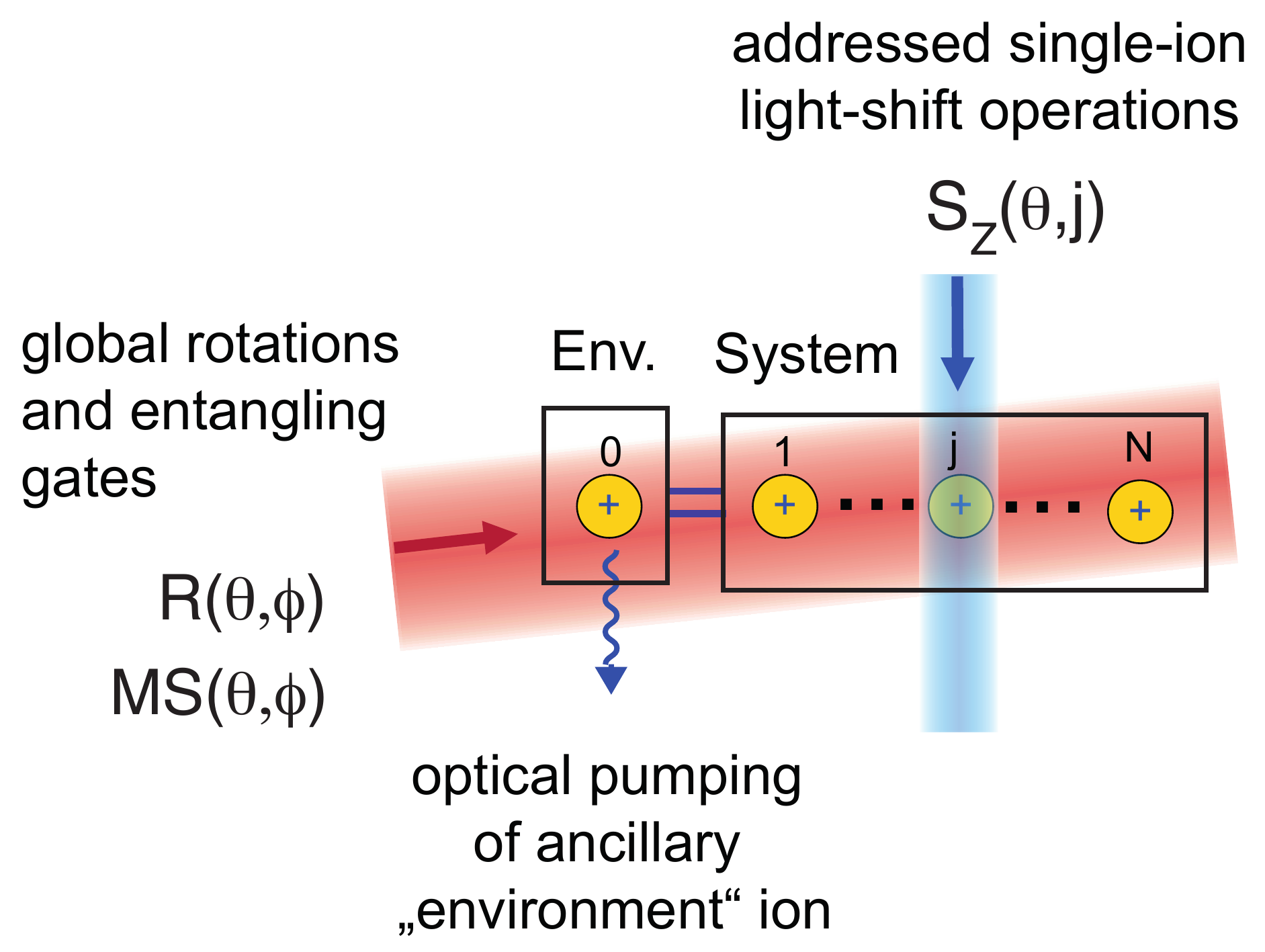}
\caption{Illustration of the geometry of the laser-beams used for
  qubit manipulation. A global beam illuminating the ion string homogeneously is used to implement collective rotations $R(\theta, \phi)$ and multi-ion M\o lmer-S\o rensen-type entangling gates $MS(\theta, \phi)$, whereas an addressed beam enables the realization of single-qubit rotations $S_Z(\theta, j)$. An addressed optical pumping technique allows one to incoherently reinitialize the state of an ancillary qubit, which plays the role of a tailored environment. See text for more details.}
\label{fig:ion_trap_tools}
\end{figure}
The qubit is encoded in the $4S_{1/2}(m_j=-1/2)$ and $3D_{5/2}(m_j=-1/2)$ electronic states of the
$^{40}$Ca$^+$ ion and is manipulated by precisely timed light pulses
on resonance or near resonant with the optical transition. The laser
light can be applied from two different directions as depicted in
figure~\ref{fig:ion_trap_tools}, where one beam illuminates the entire
ion string homogeneously and the second beam is able to address each ion
individually~\cite{fsk-decoherence}.

The set of coherent gates consists of collective single-qubit rotations, addressed single-qubit
gates and collective entangling gates. Collective single-qubit
rotations are implemented by the globally applied laser beam, resonant
with the qubit transition, realizing the unitary
\begin{equation}
\label{eq:rotations}
R(\theta, \phi)  = \exp \left(- \frac{i \theta \pi }{ 2} S_\phi \right), 
\end{equation}
with 
$S_\phi = \sum_{i=0}^N \sigma_i^\phi$ a collective spin operator, and 
$$
\sigma_i^\phi =  \sigma_i^x \cos (\phi \cdot \pi) + \sigma_i^y
\sin (\phi  \cdot \pi)
$$
being a linear combination of the single-qubit Pauli matrices $\sigma_i^{x}$ and $\sigma_i^{z}$, acting
on qubit $i$. The rotation angle $\theta$ is determined by $\theta = \Omega \, \tau / \pi$, which can be controlled by the Rabi frequency $\Omega$ and the pulse duration $\tau$. In this notation a complete $\pi$-flop inverting the electronic population of the logical states corresponds to a rotation angle $\theta=1$. On the other hand, tuning the phase $\phi$ of the global laser beam allows one to control the rotation axis lying in the $x$-$y$-plane of the Bloch sphere, around which each of the qubits is rotated. Addressed single-qubit operations
\begin{equation}
S_z(\theta, i) = \exp \left( - \frac{i \theta \pi }{2} \sigma_i^z \right)
\end{equation}
around the $z$-axis are realized by shining in laser light near resonant with the qubit transition that induces an intensity-dependent AC-Stark shift $\Delta_{AC}$. Again, the rotation angle $\theta$ is determined by the pulse
length $\tau$, $\theta = \Delta_{AC} \, \tau / \pi$, and a $\pi$-pulse corresponds to $\theta =1$. Finally, collective entangling operations are implemented by a bi-chromatic, globally 
applied laser field, which effectively realizes two-body M{\o}lmer-S{\o}rensen (MS) type interactions,
\begin{equation}
\label{eq:MS_gate}
MS(\theta, \phi) = \exp \left(  -\frac{i \theta \pi}{4} S_\phi^2 \right) = \exp \left( -\frac{i \theta \pi}{2} \sum_{i > j} \sigma_i^\phi \sigma_j^\phi \right)
\end{equation}
between all pairs $i$ and $j$ of the ion chain ($i,j=0,1,\ldots
N$)~\cite{molmer-prl-82-1835,roos-njp-10-013002}. Again, the angle
$\phi$ allows one to control whether $\sigma_i^x\sigma_j^x$ (for $\phi = 0$),
$\sigma_i^y\sigma_j^y$ (for $\phi = 1/2$) or interactions $\sigma_i^\phi \sigma_j^\phi$ corresponding to
any other axis $\sigma^\phi$ in the $x$-$y$-plane are realized. In
this notation, the angle $\theta = 1/2$ corresponds to a
"fully-entangling" MS gate, i.e.~a unitary which maps the
computational basis states of $N+1$ ions onto multi-particle entangled
states, which are (up to local rotations) equivalent to $N+1$-qubit
GHZ states. Altogether, operations in Eq. (\ref{eq:rotations}) to (\ref{eq:MS_gate}) form
a universal set of gates, enabling the implementation of arbitrary unitaries on
any subset of ions~\cite{nebendahl-pra-79-012312}.

\subsection{Numerical optimization of gate sequences}
\label{sec:numerical_optimization}
Any unitary operation required for the implementation of dissipative and coherent maps, as well the error detection and reduction protocols, 
needs to be decomposed into a sequence of available operations. As discussed in more detail below, such decompositions can be constructed systematically \cite{mueller-njp-13-085007}. However, as such decompositions are in many cases not optimal in terms of the number of required gate operations, it is convenient to resort to a numerical optimal control algorithm~\cite{nebendahl-pra-79-012312} to search for optimized sequence decompositions involving less gates. Whereas the numerical optimization algorithm becomes inefficient for general unitary operations acting on a large number of qubits, it is well-suited for the optimization of unitaries which act only on a small subset of ions (such as 2+1 ions in the implementation of an elementary dissipative Kraus map), independently of the total system size.

Numerically optimized pulse sequences may include global AC-Stark pulses and MS gates with negative rotation angles $\theta < 0$, which are not directly contained in the available gate set discussed in Sec. \ref{sec:SI_coherent_gates}. However, as any
collective rotation around the $z$-axis of the Bloch sphere can be
interpreted as a re-definition of the $x$- and $y$-axes, a
global AC-Stark pulse can be omitted if the phases $\phi$ of the following
resonant operations are properly adjusted. Regarding MS gates with $\theta < 0$, these can be implemented by MS gates with positive rotation angles, as $MS (- \theta, \phi) \equiv MS(1-\theta, \phi)$ up to local rotations (see Eq.~(9) in Ref.~\cite{mueller-njp-13-085007}). 

\subsection{Spectroscopic decoupling of ions}
\label{sec:decoupling}
Despite the globally applied beams for the collective rotations and MS
gates, operations on subsets of ions can be realized by
spectroscopically decoupling ions not involved in the realization of a
certain Kraus map from the dynamics. This is realized by coherently
transferring ions to electronic states, where they do not
couple to the globally applied light fields to a very good
approximation. This decoupling technique enables the use of optimized
sequences for the realization of Kraus maps on a small number of
sites, independently of which subset of ions is currently involved in
the map, and independently of the system size, i.e.~the total number of
sites.

The full decoupling sequence consists of the following parts as also
outlined in Fig.~2 in the main-text: (i) First, the population
is transferred from the qubit state $|S\rangle = 4S_{1/2}(m_f=-1/2) = \ket{1} = \ket{\uparrow}$ to
the $|D'\rangle = 3D_{5/2}(m_f=-5/2)$ state.  (ii) Then, the
population from the other qubit state $|D\rangle = 3D_{5/2}(m_f=-1/2) = \ket{0} = \ket{\downarrow}$
is transferred via $|S'\rangle = 4S_{1/2}(m_f=+1/2)$ to the $|D''\rangle
= 3D_{5/2}(m_f=-3/2)$ state. The required pulses for decoupling two
ions are shown in Table~\ref{tab:Decouple}. 
Bringing the population from the decoupled states
back to the original qubit states is realized by implementing the
described sequence in reverse order.

\begin{table}
  \centering
  \begin{tabular}{|c|c|}
    \hline
    Transition & Pulse \\
    \hline
$|S\rangle \rightarrow |D'\rangle $ & $R(0.5,0)$ \\
$|S\rangle \rightarrow |D'\rangle $ & $S_z(1,i)$ \\
$|S\rangle \rightarrow |D'\rangle $ & $S_z(1,j)$ \\
$|S\rangle \rightarrow |D'\rangle $ & $R(0.5,1)$ \\
\hline
$|D\rangle \rightarrow |S'\rangle $ & $R(0.5,0)$ \\
$|D\rangle \rightarrow |S'\rangle $ & $S_z(1,i)$ \\
$|D\rangle \rightarrow |S'\rangle $ & $S_z(1,j)$ \\
$|D\rangle \rightarrow |S'\rangle $ & $R(0.5,1)$ \\ \hline
$|S'\rangle \rightarrow |D''\rangle $ & $R(1,0)$ \\
\hline

  \end{tabular}
  \caption{Pulse sequence for spectroscopic decoupling of ions $i$ and $j$ by coherently transferring their quantum information from the qubit states $|S\rangle$ and $\ket{D}$ to the states $|D''\rangle$ and $|D''\rangle$. The gates listed in the three blocks of the table realize the pulses (i) - (iii) shown in Fig.~2b of the main text.}
  \label{tab:Decouple}
\end{table}

The decoupling technique introduces additional errors that are not
included in the theoretical error model. A rigorous treatment of these
errors is cumbersome since it cannot be modeled in a qubit system
anymore, but involves the full electronic substructure of the ion.
However, the effect of the decoupling process in the computational
basis can be characterized by quantum process tomography. We found a
process fidelity of 94(2)\% for decoupling a qubit and bringing it
back to the original states. Next, we proved that the decoupled qubit
is indeed to a high degree not affected by the manipulation pulses. We
checked this in a system of 3+1 ions, where we first decoupled a
single qubit, then applied the pulses as required for a single
elementary dissipative Kraus map on the remaining two system ions and
the ancilla ion, and finally transferred the decoupled qubit back to
the original qubit state. Due to residual far off-resonant coupling to
transitions coupling different Zeeman sublevels, the pulses
implementing the dissipative Kraus map induce a deterministic AC-Stark
shift on the decoupled ion. This Stark shift is measured with a
Ramsey-type experiment, and its compensation is performed with the
final two pulses in the sequence as shown in
Table~\ref{tab:Decouple}. Quantum process tomography on the decoupled
qubit, where the systematic Stark shift has been compensated, yields a
process fidelity of 93(2)\%. Thus we can conclude that the pulses
corresponding to the Kraus map implementation do not affect the
decoupled qubit significantly, and that the dominant errors result
from laser intensity fluctuations in the decoupling pulses themselves.

\subsection{Incoherent reinitialization of individual ions}
\label{sec:SI_reinitialization}
The implementation of an elementary dissipative Kraus map is completed by an incoherent reset of the ancillary ion to its initial state $|S\rangle$, see step (iii) in Fig.~2c in the main text. This reinitialization is realized by an optical pumping technique: First, an addressed pulse is applied to the ancillary ion to transfer the population present in the $|D\rangle$ state to the 
$|S'\rangle$ state. Then $\sigma^-$ polarized light is applied to the
entire ion string performing optical pumping from $|S'\rangle$ towards $|S\rangle$ via the short-lived $4^2P_{1/2}$ state. This
procedure does not affect the information in the system ions encoded
in the original qubit states $|S\rangle$ and $|D\rangle$, as the light couples only to the
$|S'\rangle$ level \cite{barreiro-nature-470-486}. The required operations are shown Table~\ref{tab:Reset}

\begin{table}
  \centering
  \begin{tabular}{|c|c|}
    \hline
    Transition & Pulse \\
    \hline
$|S'\rangle \rightarrow |D\rangle $ & $R(0.5,0)$ \\
$|S'\rangle \rightarrow |D\rangle $ & $S_z(1,j)$ \\
$|S'\rangle \rightarrow |D\rangle $ & $R(0.5,1)$ \\
\hline
$|S'\rangle \rightarrow |P\rangle $ & $\sigma^-$ repump \\
\hline
  \end{tabular}
  \caption{Pulse sequence for the individual reset of qubit~$j$.}
  \label{tab:Reset}
\end{table}

\section{Engineerging of dissipative and Hamiltonian dynamical maps}
\label{sec:engineering} 
In this appendix we provide details on the engineering and the specifics of the circuit-based experimental implementation of elementary dissipative and Hamiltonian dynamical maps. 

\subsection{Action of the dissipative Kraus maps} 
\begin{figure}
\center
\includegraphics[angle=0,width=0.6\columnwidth]{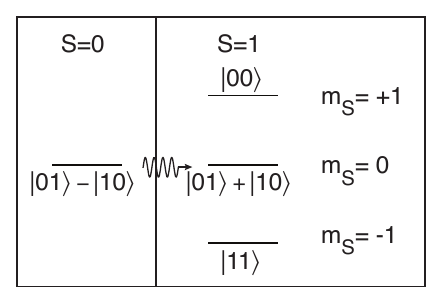}
\caption{Schematics of the action of the two-spin operator (\ref{eq:supp_jump}) in the local Hilbert space of the two spins $i$ and $i+1$, which is spanned by the singlet state (left sector of the Hilbert space) and the triplet states (right sector). The operator $c_{i}$ converts the singlet state $\frac{1}{\sqrt{2}} (\ket{01} - \ket{10})$ into the triplet $m_S=0$ state $\frac{1}{\sqrt{2}} (\ket{01} + \ket{10})$. All triplet states are dark states and left invariant.}
\label{fig:singlet_to_triplet_pumping}
\end{figure}

The elementary dissipative Kraus maps
\begin{eqnarray}
\label{eq:supp_dynmap}
\rho \mapsto E_{i,1} \rho  E_{i,1}^{\dag} + E_{i,2} \rho  E_{i,2}^{\dag}
\end{eqnarray}
with 
\begin{equation}
\label{eq:supp_Ei}
E_{i,1} = c_i \qquad \text{and}\qquad  E_{i,2} = 1 - c_i^\dagger c_i,
\end{equation}
are generated by the operators $c_i$ (cf.~Eq.~(2) in the main text), as given by
\begin{equation}
\label{eq:supp_jump}
c_i = (\sigma_i^+ + \sigma_{i+1}^+ )  (\sigma_i^- - \sigma_{i+1}^- ).
\end{equation}
These operators act bi-locally, i.e.~involve two neighboring spins $i$ and $i+1$, whereas the other spins are spectators. It is instructive to examine their action on the basis states of the local Hilbert space of the two qubits $i$ and $i+1$, which is spanned by the singlet and triplet states of the total spin $S_{i,i+1}^2$ of the two spin-1/2 particles:
$$
S_{i,i+1}^2 \frac{1}{\sqrt{2}} (\ket{01} - \ket{10}) = 0
$$
and
\begin{eqnarray}
S_{i,i+1}^2 (\ket{00}, \, \frac{1}{\sqrt{2}} (\ket{01} + \ket{10}), \, \ket{11}) \nonumber \\
\qquad = 2 \, (\ket{00}, \,  \frac{1}{\sqrt{2}} (\ket{01} + \ket{10}), \, \ket{11}). \nonumber
\end{eqnarray}
Here, $\mathbf{S}_{i,i+1} = \mathbf{S}_i + \mathbf{S}_{i+1}$ with $\mathbf{S}_i = \frac{1}{2}\mathbf{\sigma}_i = \frac 12 (\sigma_i^x, \sigma_i^y, \sigma_i^z)^T$ and 
$$
\mathbf{S}_{i,i+1}^2 = (\mathbf{S}_i + \mathbf{S}_{i+1})^2 = \frac{3}{2} + \frac{1}{2} (\sigma_i^x \sigma_{i+1}^x + \sigma_i^y \sigma_{i+1}^y +\sigma_i^z \sigma_{i+1}^z).
$$
For simplicity of the notation, we skip the spin indices $i$ and $i+1$ for the states and use the short-hand notation $\ket{00} = \ket{0}_i \otimes \ket{0}_{i+1}$, etc. As illustrated in Fig.~\ref{fig:singlet_to_triplet_pumping}, the operators $c_i$ induce pumping from the singlet into the triplet $m_S = 0$ state, 
$$
c_{i}  \frac{1}{\sqrt{2}} (\ket{01} - \ket{10}) = \frac{1}{\sqrt{2}} (\ket{01} + \ket{10}),
$$
whereas all triplet states are dark states, 
$$
c_{i} \ket{00} = c_{i} \frac{1}{\sqrt{2}} (\ket{01} + \ket{10}) = c_{i} \ket{11} = 0.
$$
As shown in Fig.~1b of the main text, under this dissipative dynamics a single spin excitation (or hardcore boson) is symmetrically delocalized over the two sites, whereas the states of two ($\ket{11}$) or zero ($\ket{00}$, "vacuum") spin excitations (or hardcore bosons) are left unchanged. 

\subsection{Circuit-based implementation of elementary dissipative Kraus maps}
\label{sec:supp_diss_Kraus_maps}
Open-system dynamics according to the elementary dissipative Kraus maps (\ref{eq:supp_dynmap}) can be realized in a ''digital" way, by using  quantum simulation tools for open systems, which have been previously developed and demonstrated experimentally in the context of dissipative preparation of Bell and multi-qubit stabilizer states~\cite{barreiro-nature-470-486}. The key idea in engineering the two-spin dissipative dynamics according to Eq.~(\ref{eq:supp_dynmap}) is to combine the experimentally available gates (see Sec.~\ref{sec:SI_coherent_gates} below for details) with optical pumping on an additional ancillary qubit (see Sec.~\ref{sec:SI_reinitialization}), which plays the role of a tailored environment.
 
\textit{General engineering strategy} -- The observation that the singlet state is dissipatively converted into the $m_S=0$ triplet state suggests the following gate-based implementation via a four-step process (shown in Figure~\ref{fig:generic_circuits}a), which involves a circuit of unitaries applied to the qubits $i$, $i+1$ (steps (i) to (iii)), followed by the incoherent reset of the ancilla qubit (step (iv)):
\begin{figure}
\center
\includegraphics[angle=0,width=1\columnwidth]{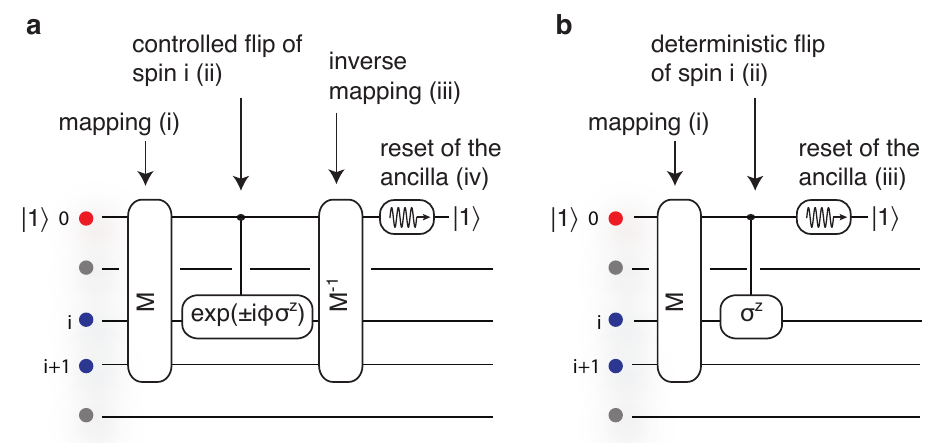}
\caption{Quantum circuits for the realization of elementary dissipative Kraus maps (\ref{eq:supp_dynmap}) on spins $i$ and $i+1$. \textbf{a,} First, the information whether the two system spin-1/2 particles $i$ and $i+1$ are in the singlet or triplet subspace of the two-spin Hilbert space (see Fig.~\ref{fig:singlet_to_triplet_pumping}) is coherently mapped onto the logical states $\ket{0}$ and $\ket{1}$ of an ancilla qubit which is initially in $\ket{1}$. (ii) Next, a two-qubit gate is applied, which realizes an effective spin flip on one of the system spins, and thereby conditionally on the state of the ancilla converts the singlet into the triplet $m_S=0$ state: $\frac{1}{\sqrt{2}} (\ket{01} - \ket{10}) \rightarrow \frac{1}{\sqrt{2}} (\ket{01} + \ket{10})$. After inverting the mapping (iii), the ancilla ion is optically pumped backed to its initial state $\ket{1}$ (iv). This last steps renders the dynamics irreversible and provides the dissipative ingredient to extract entropy from the two system spins. The probability of the conversion from the singlet to the triplet state is controlled by the angle $\phi$ appearing in the two-qubit gate, and it is given by $p = \sin^2 \phi$. In the limit of $\phi \ll \pi/2$, the conversion from the singlet into the triplet state takes place only with a small probability, and the general dissipative Kraus map describing the pumping process reduces to the master equation (\ref{eq:master_equation}). \textbf{b,} In this work we are interested in deterministic pumping from the singlet into the triplet subspace, i.e.~in the case of probability $p=1$ for $\phi = \pi/2$. In this situation, the circuit simplifies as the inverse mapping to partially disentangle the system spins from the ancilla is not required, and the circuit simplifies to the three-step process, which is also shown in Fig.~2c of the main text.} 
\label{fig:generic_circuits}
\end{figure}

(i) First, a unitary $M$, acting on the two system spins $i$ and $i+1$ and the ancillary qubit, coherently maps the binary information whether the two system spins are in the singlet or triplet subspace onto the two logical states $\ket{0}$ and $\ket{1}$ of the ancillary qubit. This is achieved by the unitary
\begin{eqnarray}
\label{eq:M}
M(\theta) & = & \exp\left( -\frac{i\theta}{2} \left( \mathbf{S}_{i,i+1}^2-2 \right) \otimes \sigma_0^x \right) \\
& = & \exp \left( i\theta P_{i,i+1} \otimes \sigma_0^x \right) \nonumber \\
& = & \cos(\theta P_{i,i+1}) \otimes 1_0- i \sin(\theta P_{i,i+1}) \otimes \sigma_0^x \nonumber
\end{eqnarray}
for which we choose $\theta = \pi/2$ so that the unitary reduces to
\begin{equation}
\label{eq:mapping}
M(\pi/2) = (1-P_{i,i+1}) \otimes 1_0 - i \,P_{i,i+1} \otimes \sigma_0^x.
\end{equation}
The unitary $M$ is constructed in a way that the state of the ancilla qubit is rotated conditional on the state of the two system spins, and that the angle of this rotation depends on the operator $(\mathbf{S}_{i,i+1}^2-2)$ which acts on the two system spins $i$ and $i+1$. Here, $P_{i,i+1} = c_i^\dagger c_i$ is the projector onto the singlet subspace, $(1-P_{i,i+1})$ its orthogonal complement, and we have used $\mathbf{S}_{i,i+1}^2 = 2 (1-P_{i,i+1})$. Under the unitary $M(\pi/2)$, the ancilla qubit is rotated from its initial state $\ket{1}$ into $\ket{0}$ if and only if the two systems spins are in the singlet state.

(ii) Next, the transfer of the system qubits from the singlet to the triplet subspace is enabled by a two-qubit gate, which acts on the ancilla qubit and the system qubit $i$. It reads
\begin{align}
\label{eq:correction_gate}
C(\phi) = & \exp \left( \pm \frac{1}{2} \left( 1 - \sigma_0^z \right) \otimes i \phi \sigma_i^z\right) \\
= &  \ket{1}\bra{1}_0 \otimes 1_i + \ket{0}\bra{0}_0 \otimes \exp(\pm i \phi \sigma_i^z). \nonumber
\end{align}
This controlled-operation applies $\sigma^z$ to the $i$-th system qubit, and thus with probability $p = \sin^2 \phi$ converts the singlet into the triplet state: $\sigma_i^z \frac{1}{\sqrt{2}} (\ket{01} - \ket{10}) = \frac{1}{\sqrt{2}} (\ket{01} + \ket{10})$. Here, the ancilla qubit acts as a \textit{quantum controller} \cite{lloyd-pra-65-010101,Baggio12}, which is not observed but controls the coherent feedback which is applied to the system qubits. From Eq.~(\ref{eq:correction_gate}) it is clear that if the ancilla is in state $\ket{1}$, corresponding to the case of the two system spins residing in the triplet subspace, no quantum feedback is applied and the state of the system spins remains  unchanged. Due to the previous mapping $M$, this assures that the triplet subspace is left invariant. 

(iii)
The initial mapping (i)  is inverted by applying the inverse unitary $M^\dagger (\pi/2)$. For the system being in the target triplet space from the outset, this re-installs the conditions before step (i). For the special case of deterministic pumping ($p=1$), as experimentally realized in this work, the inverse unitary can be omitted (see also Figure~\ref{fig:generic_circuits}b).

(iv)
After these unitary steps, the ancillary qubit remains in general entangled with the two system spins $i$ and $i+1$. It is finally reset to its initial state $\ket{1}$ by optical pumping: this is the physical dissipative mechanism, which renders the dynamics irreversible and allows to extract entropy from the system spins. This refreshes the ancilla qubit and prepares it in the known pure state $\ket{1}$, so that it can be used for the implementation of subsequent dissipative Kraus maps.

For an initially uncorrelated state of the ancilla qubit and the system spins, $\ket{1}\bra{1}_0 \otimes \rho $, the resulting dynamics for the system spins (described by the reduced density matrix $\rho$ of the system spins) is obtained by applying the combined unitary $U = M^\dagger(\pi/2) \, C(\phi) \,M(\pi/2)$ and tracing over the ancilla qubit's degrees of freedom, 
\begin{equation}
\label{eq:map_and_trace}
\rho \mapsto \mathrm{tr}_{0} \left\{ U (\ket{1}\bra{1}_0 \otimes \rho) U^\dagger\right\}.
\end{equation}
Straightforward algebra yields the Kraus map $\rho \mapsto \sum_{k=1,2} E_{i,k} \rho E_{i,k}^\dagger$ with the operation elements 
\begin{eqnarray}
E_{i,1} & = \sin \phi \,c_i \qquad \text{and} \qquad E_{i,2} = 1 + (\cos \phi - 1) \,c_i^\dagger c_i. \nonumber
\end{eqnarray}
One the one hand, for $\phi = \pi/2$ the operation elements reduce to Eq.~(\ref{eq:supp_Ei}) and correspond to deterministic pumping ($p=1$) from the singlet into the triplet states, which is the scenario realized in this work. On the other hand, for $\phi \ll \pi/2$ one can expand the operation elements of the Kraus map and recovers the master equation limit
\begin{equation}
\label{eq:master_equation}
\rho \mapsto \rho + \phi^2 \left(c_i \rho c_i^\dagger - \frac{1}{2} \left\{ c_i^\dagger c_i, \rho \right\} \right) + O(\phi^4).
\end{equation}

\textit{Additional remarks} -- We note that our implementation corresponds to an \textit{open-loop} control scenario, where the ancillary qubit remains unobserved during the simulation. However, we remark that it is possible to measure the state of the ancillary qubit in the computational basis by an addressed fluorescence measurement, before it is reset to its initial state by optical pumping~\cite{barreiro-nature-470-486}. Such measurement reveals whether under the application a particular dissipative Kraus map the two system spins have undergone a collective two-spin quantum jump from the singlet into the triplet state or not. This information yields "in-situ" information about the many-body system's dynamics along a particular trajectory. The temporal statistics of quantum jumps in open, driven many-body quantum spin systems contains valuable information about dynamical phase transitions \cite{garrahan-prl-104-160601,ates-pra-85-043620}. In the present work we do not further explore this possibility. We remark that the fluorescence measurement of the ancillary ion, which is associated to the scattering of many photons and non-negligible heating of the vibrational modes, can be combined with subsequent laser re-cooling of the relevant vibrational modes via the ancillary ion, as demonstrated by~\cite{Schindler12}. This allows one to perform such quantum jump measurements and to afterwards re-initialize the \textit{external} degrees of freedom, thereby enabling that in the same experimental run further gate operations required for the implementation of subsequent Kraus maps can be still applied with high fidelity. 

\textit{Specific experimental implementation} -- For the implementation of the elementary dissipative Kraus maps in the ion-trap simulator, the unitary operations $M(\pi/2)$ and $C(\pi/2)$ (see Eqs.~(\ref{eq:M}) and (\ref{eq:correction_gate})) have to be decomposed into a quantum circuit of available gates: The mapping $M$ can be written as a product of three unitaries involving 3-body spin interaction terms,
\begin{eqnarray}
M ( \pi/2) = \exp\left( -\frac{i \pi}{4} \sigma_0^x (\mathbf{S}_{i,i+1}^2-2)\right) \hspace{2cm} \\
= \exp\left( + \frac{i \pi}{8} \sigma_0^x\right) \exp\left( - \frac{i \pi}{8} \sigma_0^x \sigma_i^x \sigma_{i+1}^x \right) \hspace{14mm}\nonumber \\
\times \exp\left( - \frac{i \pi}{8} \sigma_0^x \sigma_i^y \sigma_{i+1}^y \right)
\exp\left( - \frac{i \pi}{8} \sigma_0^x \sigma_i^z \sigma_{i+1}^z \right). \hspace{3mm} \nonumber
\label{eq:U_M_decomposition}
\end{eqnarray}
As discussed in Ref.~\cite{mueller-njp-13-085007} each of the three-body unitaries can be realized by two "fully-entangling" MS gates in combination with single-qubit rotations, such that $M$ could be implemented by a quantum circuit involving six MS gates and a number of single-qubit gates to rotate the system spins between the $x$, $y$ and $z$-bases between the different unitaries. 

Similarly, the two-qubit controlled operation $C(\pi/2)$ (see Eq.~(\ref{eq:correction_gate})) is up to local unitaries equivalent to one "fully-entangling" two-ion MS gate \cite{mueller-njp-13-085007}, 
$$C(\pi/2)  \sim \exp(- i (\pi/4) \sigma_0^x\sigma_i^x),
$$
acting on the ancilla qubit and the system spin $i$. In the experiment, we did not use such a systematically constructed circuit decomposition, but instead resorted to the numerical optimization algorithm described in Sec.~\ref{sec:numerical_optimization} to further reduce the complexity of the quantum circuit: The gates of the experimentally employed sequence decomposition for the implementation of the unitaries of one elementary dissipative Kraus map are listed in Table~\ref{tab:SingleSeq}.
\begin{table}[h]
  \centering
  \begin{tabular}{|c|c||c|c|}
    \hline
    Number & Pulse & Number & Pulse\\
    \hline
1 & $S_z(1.5, 0)$  & 11 & $MS(0.5, 1.5)$  \\ 
2 & $R(1.5, 1.0)$  &  12 & $S_z(1.75, 2)$  \\ 
3 & $MS(0.25, 1.0)$  & 13 & $MS(0.5, 2.25)$  \\ 
4 & $S_z(1.0, 1)$  & 14 & $R(0.5, 1.75)$  \\ 
5 & $MS(0.875, 1.0)$  & 15 & $R(0.5, 2.25)$  \\ 
6 & $S_z(1.0, 2)$  & 16 & $MS(0.25, 2.25)$  \\ 
7 & $MS(0.125, 1.0)$  & 17 & $S_z(1.5, 2)$  \\ 
8 & $S_z(1.5, 2)$  & 18 & $S_z(1.0, 1)$  \\ 
9 & $R(0.5, 0.5)$  &19 & $R(0.5, 2.25)$  \\ 
10 & $S_z(0.5, 2)$  & &\\
    \hline
  \end{tabular}
  \caption{Pulse sequence for the implementation of a single elementary dissipative map. The necessary operations for the reset step are not shown. }
  \label{tab:SingleSeq}
\end{table}

\subsection{Engineering Hamiltonian maps for competing coherent interactions}
\label{sec:maps}
Competing Hamiltonian dynamical maps are realized according to the dimensionless many-body spin Hamiltonian (cf.~Section~\ref{sec:main_competing} of the main text)
\begin{equation}
H = \sum_{i=1}^{N-1} H_i  := \sum_{i=1}^{N-1} (1+\sigma^z_i) (1+\sigma^z_{i+1})/4,
\end{equation}
where the two-body terms correspond to interactions between spin excitations (or hardcore bosons) located on neighboring sites $i$ and $i+1$. The Hamiltonian dynamical maps $U_{i,i+1} = \exp (-i \phi H_i)$ acting on spins $i$ and $i+1$ are up to local rotations equivalent to two-spin MS interactions, $\exp (- i \phi \,\sigma_i^x \sigma_{i+1}^x/4)$. The implementation of the competing Hamiltonian dynamical maps can be realized with two distinct approaches: (i) The elementary two-spin Hamiltonian maps can be implemented sequentially, in analogy to the sequential implementation of the elementary dissipative maps. (ii) Alternatively, since the elementary Hamiltonian maps mutually commute, they can be implemented by a single global unitary operation, acting directly on the entire register of system spins. The unitary according to a sum of pairwise interactions between \textit{neighboring} spins can for instance be built up from MS gates, which involve pairwise interactions between \textit{all} pairs of spins, by means of refocusing techniques \cite{mueller-njp-13-085007}. Although one could again try to find numerically optimized sequences of gates for the implementation of the composite Hamiltonian map, we note that such optimization for the unitary acting on the entire register of system spins would have to be done for each system size, and becomes inefficient for increasing system sizes. It is thus advisable to exploit the symmetries of the MS interactions, and to systematically construct sequence decompositions. 

\textit{Specific experimental implementation} --  For instance, in the case of 3 system spins (with open boundary conditions), this operation can be decomposed into the following unitaries according to single- and two-body interactions
\begin{align}
\label{eq:U_competition}
U_\mathrm{comp} = & \exp \left(- \frac{i \phi}{4} (\sigma_1^z \sigma_2^z +\sigma_2^z \sigma_3^z ) \right)\\
\times & \, \exp \left(- \frac{i \phi}{4} (\sigma_1^z + \sigma_2^z +\sigma_3^z) \right) 
 \, \exp \left(- \frac{i \phi}{4} \sigma_2^z \right). \nonumber
\end{align}
As the ancilla qubit is not required to realize the unitary map, and must not be entangled with the system spins during the operation, it is spectroscopically decoupled during the application of the gate sequence. The experimental sequences used for the implementation of competing Hamiltonian maps for 3 system spins (3+1 ions) and 4 system spins (4+1 ions) are shown in Tables~\ref{tab:Comp3} and~\ref{tab:Comp4}, respectively.

\begin{table}[h]
  \centering
  \begin{tabular}{|c|c|}
    \hline
    Number & Pulse \\
    \hline
1 & $S_z(1.5, 1)$  \\ 
2 & $R(0.5, 0.0)$  \\ 
3 & $MS(1 - 0.25 \, k, 0.5)$  \\ 
4 & $S_z(1.0 , 1)$  \\ 
5 & $MS(0.25 \, k, 0.5)$  \\ 
6 & $S_z(1.0, 1)$  \\ 
7 & $R(0.5, 1.0)$  \\ 
    \hline
  \end{tabular}
  \caption{Pulse sequence for the implementation of a composite Hamiltonian map in a 3-spin system, according to Eq.~(\ref{eq:U_competition}). The Hamiltonian strength is controlled by the parameter $k$, with experimentally implemented values $k \in \{1,0.5\}$, corresponding to $\phi \in \{ \pi/2, \pi/4 \}$.}
  \label{tab:Comp3}
\end{table}

\begin{table}[h]
  \centering
  \begin{tabular}{|c|c|}
    \hline
    Number & Pulse \\
    \hline
 & Act on qubit 2 and 3 \\ \hline
1 & $R(0.5, 0.5)$  \\ 
2 & $MS(0.5, 1.0)$  \\ 
3 & $R(0.5, 1.5)$  \\  \hline
 & Act on qubit 1,2,3,4 \\ \hline
4 & $R(0.5, 0.5)$  \\ 
5 & $MS(0.25, 1.0)$  \\ 
6 & $S_z(1.0, 3)$  \\ 
7 & $S_z(1.0, 2)$  \\ 
8 & $MS(0.25, 1.0)$  \\ 
9 & $S_z(1.5, 3)$  \\ 
10 & $S_z(1.5, 2)$  \\ 
11 & $R(0.5, 1.5)$  \\ 
    \hline
  \end{tabular}
  \caption{Pulse sequence for the implementation of a composite Hamiltonian map in a 4-spin system, for strong competing interactions corresponding to an angle $\phi = \pi/2$.}
  \label{tab:Comp4}
\end{table}

\section{Experimental characterization of an elementary dissipative map and noise model}
\label{sec:exp_characterization} 

\subsection{Modeling an imperfect elementary dissipative map}
\label{sec:modeling_the_process}

In the following, we will introduce a theoretical model of the
elementary pumping step in the presence of experimental noise.  A
single elementary dissipative map acting on two system ions can be
ideally described by the two-qubit process matrix $\chi_{id}$
which can be straightforwardly calculated from the Kraus map
(\ref{eq:supp_dynmap}) with generating operators as defined in
Eq.~(\ref{eq:supp_Ei}) \cite{nielsen-book}. In order to describe the implementation of the
elementary dissipative map on the three system qubits (two "active"
and one spectroscopically decoupled ion), we will assume a process
consisting of the modeled elementary map on the two active system
qubits and an identity process on the third qubit. The noise affecting
the elementary map is modeled by two independent depolarizing channels
acting on each of the two "active" system qubits where the fully
depolarizing channel on a single qubit $i$ can be written as the Kraus map \cite{nielsen-book}
\[
\rho \mapsto \mathcal{E}_{dep}^{(i)}(\rho) = \frac{1}{4}\left( \rho +  \sigma_i^{x} \rho \sigma_i^x +\sigma_i^{y} \rho \sigma_i^y+\sigma_i^{z } \rho \sigma_i^z \right).
\]
The physical noise acting on the register during the individual gates
is certainly more complex than this, but the effect of noise on the
outcome of any complex algorithm can be approximated by depolarizing
noise regardless of the specific characteristics of the noise
\cite{nielsen-book}. We model a noisy elementary dissipative map with
the concatenation of two depolarizing channels each acting on one of
the system qubits $\rho \mapsto \Pi(\rho) =
\mathcal{E}_{dep}^{(i+1)}(\mathcal{E}_{dep}^{(i)}(\rho))$ with process matrix $\chi_\Pi$ as
\[
\chi_{diss}(\epsilon)=(1-\epsilon)\chi_{id}+\epsilon \, \chi_\Pi.
\]
In the limit $\epsilon \rightarrow 0$, the imperfect process $\chi_{diss}$ reduces to the ideal two-qubit process $\chi_{id}$, where in the extreme opposite limit of $\epsilon \rightarrow 1$ depolarizing noise completely dominates and overwrites any effect of the desired engineered dissipative process $\chi_{id}$. One can now adjust the parameter $\epsilon$ of this model to the
obtained data by a numerical optimization. For this, we
maximized the overlap between the expected output state after a single
elementary map $\rho_\epsilon$ with noise strength $\epsilon$ with the
actual measured state $\rho_\mathrm{exp}$,
\[
\arg_\epsilon \max \mathcal{F}(\rho_\mathrm{exp}, \rho_\epsilon).
\]
where the Uhlmann fidelity $\mathcal{F}(\rho_1,\rho_2)$ for the comparison of two density matrices $\rho_1$, $\rho_2$ is used
\cite{gilchrist-pra-71-062310}.  We find an optimum for a noise parameter of
$\epsilon=0.27$.

\subsection{Implementation and analysis of an elementary dissipative
  map}
\label{sec:implementationmap}
Here, we will provide a more detailed analysis of the specific
implementation of a single dissipative step. During the realization of
an elementary dissipative Kraus map gates act on the two ions
encoding the system spins $i$ and $i+1$ and the ancilla ion, while all other ions are
spectroscopically decoupled. To quantitatively characterize the
implementation of the Kraus map, as realized by the gate sequence
shown in Table~\ref{tab:SingleSeq} (see also
Sec.~\ref{sec:supp_diss_Kraus_maps}), we performed a quantum process
tomography on the two system qubits. A benchmark for the performance
is given by the process fidelity with the ideal two-qubit
process. Since the ideal process is not a unitary process, the
Choi-Jamiolkowsky process fidelity is a suitable measure \cite{gilchrist-pra-71-062310}. We
find a mean state fidelity of $\mathcal{F}=68(1)\%$.

In order to identify the leading source of imperfections in the
implementation, a numerical analysis of the actual physical system is
performed. We performed a Monte Carlo simulation of a single
elementary map acting on three ions, including noise originating from:
laser frequency and intensity fluctuations, magnetic field
fluctuations, imperfect state preparation, motional heating,
spontaneous decay and crosstalk of the addressed operations.  The
noise parameters are independently measured and we find an overlap
between the numerically predicted and the measured state of
$\mathcal{F}=97 \%$. In order to determine the dominant error source,
we performed the simulation multiple times where only a single noise
source affects the evolution. The results are shown in
Table~\ref{tab:TIQC}.  From this we find that the main error source is
dephasing due to fluctuations in the laser frequency and the magnetic
field.

\begin{table}
  \centering
  \begin{tabular}{|c|c|}
    \hline
    Error source & Overlap with $\Psi_+$ \\ \hline
    All & 77 \% \\ \hline
    Addressing & 95\% \\ \hline
    Dephasing & 84 \% \\ \hline
    Intensity fluctuations & 99\% \\ \hline
    Spectator modes & 94\% \\ \hline
  \end{tabular}
  \caption{Results for the numerical simulation of a quantum 
    simulation algorithm, in order to identify the dominant experimental error source. 
    The simulation is performed multiple times with only a single active 
    error source. From the results one can infer that dephasing is 
    the dominant  error source.}
  \label{tab:TIQC}
\end{table}

\subsection{Numerical simulation and long-time dynamics under dissipative maps}
Based on the study of experimental errors of an elementary dissipative Kraus map above, we have realized a numerical study of the dissipative dynamics driving a mesoscopic spin systems ($N=10$) with initially three spin excitations present ($m_0=3$) towards the Dicke state $\ket{D(3,10)}$. The results are shown and discussed in Figure 5a of the main text. To take into account the imperfections in the implementation of the elementary maps, as discussed above, we assumed a depolarizing noise strength of $\epsilon_{diss} = 0.02$ for each of the two spins $i$ and $i+1$ involved in an elementary dissipative map $D_{i,i+1}$. Since the implementation of an elementary Hamiltonian dynamical map $U_{i,i+1}$ on two spins is less complex than the realization of a dissipative map, we assumed a depolarizing noise strength for $U_{i,i+1}$ of $\epsilon_{coh} = \epsilon_{diss}/5$. The error reduction protocol to stabilize the system within the desired $m_0=3$ excitation subspace can in practice also be only implemented with a finite accuracy. Here we assumed both for the global operation required for one spin extraction procedure and for one injection procedure that the whole spin register is exposed to a global depolarizing noise channel with a probability $\epsilon = 0.02$. 

\textit{Long-time dynamics and off-diagonal long-range order under imperfect dissipative dynamical maps} --  As discussed above and in the main text, under experimental imperfections during the pumping into Dicke states, the system initially residing in the $m_0$ excitation subspace suffers population leakage into the subspaces with $m\neq m_0$. In the long time limit, the population of the spin system is homogeneously distributed over these $m$-subspaces, each subspace populated according to its micro-canonical weight, i.e. the number of computational basis states spanning the corresponding subspace. Within each of the subspaces, not only for $m = m_0$, the dissipative maps are active and pump the population within each subspace towards the corresponding Dicke state $\ket{D(m,N)}$. As a result, for weak noise and in the long time limit, the system is driven into an incoherent mixture of Dicke states of different excitation numbers $m=0, \ldots, N$. 

More quantitatively, a normalized Dicke state with $m$ excitations for $N$ spins is given by 
\begin{equation}
\ket{D(m,N)} = (m!)^{-1}\left( \begin{array}{c} N \\ m \end{array} \right)^{-1/2} \left( S^+ \right)^m \ket{0}^{\otimes N}
\label{eq:Dicke_state_formula}
\end{equation}
with $S^+ = \sum_{k=1}^N \sigma_k^+$.
The number of micro-states of $m$ excitations on $N$ sites is $\left( \begin{array}{c} N \\ m \end{array} \right)$. Thus the incoherent mixture of Dicke states is given by
\begin{eqnarray}\label{eq:rhom}
\rho &= & \frac{1}{2^N} \sum_{m=0}^N \left( \begin{array}{c} N \\ m \end{array} \right) \ket{D(m,N)} \bra{D(m,N)} \\
&= & \frac{1}{2^N} \sum_{m=0}^N \frac{1}{(m!)^2} \left( S^+ \right)^m \ket{0}^{\otimes N} \bra{0}^{\otimes N} \left( S^- \right)^m \nonumber 
\label{eq:Dicke_mixture_expression}
\end{eqnarray}
Interestingly, this incoherent mixture of Dicke states is a state with off-diagonal order $\langle \sigma_i^+ \sigma_j^- \rangle \neq 0$, 
\begin{equation}
\label{eq:sigma_i_plus_sigma_j_minus}
\langle \sigma_i^+ \sigma_j^- \rangle = \frac{1}{N(N-1)} \left( \langle S^+ S^- \rangle - m \right),
\end{equation}
as we will show in the following. In Eq.~(\ref{eq:sigma_i_plus_sigma_j_minus}) we have made use of the symmetry of the Dicke states under permutations of the spin indices. In order to evaluate the global expectation value $\langle S^+ S^- \rangle$, we decompose it as
$\langle S^+ S^- \rangle = \langle S^- S^+ \rangle + \langle S^z \rangle$. 
Both contributions are easily obtained as
$\langle S^z \rangle = \sum_i \langle \sigma_i^z \rangle = 2m - N$,
and
$\langle S^- S^+ \rangle = (m + 1) (N-m)$.
The latter expectation value can be obtained using the normalization factors for the Dicke states $|D(m,N)\rangle$ and $|D(m+1,N)\rangle$ with $m$ and $m+1$ excitations, respectively. Adding both contributions, we obtain 
$\langle S^+ S^- \rangle = m ( N+1 - m)$.
This allows us to evaluate the expectation value $\langle \sigma_i^+ \sigma_j^- \rangle$ with respect to a pure Dicke state:
\begin{align}
\langle \sigma_i^+ \sigma_j^- \rangle_{\ket{D(m,N)}} & = \frac{1}{N(N-1)} \left( \langle S^+ S^- \rangle - m \right) \nonumber \\
& = \frac{m}{N} \left( 1 - \frac{m}{N} \right) \cdot \frac{1}{1-\frac{1}{N}}. \nonumber 
\label{eq:sigma_i_plus_sigma_j_minus_result}
\end{align}
In the thermodynamic limit, $N \rightarrow \infty$, $m/N =$ const., we have $\langle \sigma_i^+ \sigma_i^- \rangle \rightarrow m/N (1-m/N)$. This expression reflects the "particle-hole" symmetry and shows that at complete filling ($m=N$ spin excitations) or in the "vacuum" state ($m=0$) there is no off-diagonal order because spin excitations or missing spin excitations cannot be delocalized over the spin chain. At half filling $m=N/2$, where the number of micro-states is maximal, the effect of delocalization and thus the off-diagonal order is maximal. The finite-size factor $1/(1-1/N)$ is to be taken into account in small or mesoscopic spin systems, and approaches 1 in the thermodynamic limit.

From this, we can now determine the off-diagonal order of the system in the incoherent mixture of Dicke states as given by the density matrix of Eq.(\ref{eq:Dicke_mixture_expression}). Using the identities 
\begin{align}
\sum_{m=0}^N \left( \begin{array}{c} N \\ m \end{array} \right) m & = N 2^{N-1}, \nonumber \\
\sum_{m=0}^N \left( \begin{array}{c} N \\ m \end{array} \right) m^2 & = N (N+1) 2^{N-2},\nonumber
\end{align}
one obtains
\begin{equation}
\langle \sigma_i^+ \sigma_j^- \rangle_\mathrm{mixture} = \frac{1}{4}. \nonumber 
\end{equation}
As expected, this value does not depend on the initial number of excitations anymore, since this information is completely lost, once the system has diffused into the incoherent mixture of Dicke states. The off-diagonal order in the incoherent mixture of Dicke states assumes a value which is independent of the system size $N$, and remains finite in the thermodynamic limit. 

In summary, the experimental imperfections, as described by depolarizing noise and resulting in the non-conservation of the excitation-number during the simulation, lead to a strong decrease of the state overlap between the asymptotically reached many-body body state of Eq.~(\ref{eq:Dicke_mixture_expression}) and the ideal "target" dark state $\ket{D(m_0,N)}$. However, from a condensed-matter perspective one can state that the imperfections are not too harmful to the off-diagonal long-range order, measured by the two-spin correlation function $\langle \sigma_i^+ \sigma_j^- \rangle$ for $|i-j| \gg 1$ as an \textit{order parameter}. The off-diagonal order is constantly stabilized and re-built by virtue of the repeated application of the engineered (though imperfectly implemented) dissipative dynamical maps.


\section{Quantum error detection method: Post-selective QND scheme}
\label{sec:QND_postselection} 

\subsection{General idea}
\label{subsec:QND_idea}
The post-selective error detection method is based on a quantum non-demolition (QND) measurement of the total number $m$ of spin excitations present in the system at the end of the sequence of dynamical maps. Simulation outcomes, where due to experimental imperfections the ideally conserved initial spin excitation number $m_0$ has changed to a final value $m \neq m_0$ are discarded. This leads to an improvement of the simulation accuracy for longer sequences of dynamical maps, at the expense of an increased number of experimental runs to obtain the same measurement statistics. We remark that in a large system, as typical for post-selective techniques, this method becomes inefficient as the probability of remaining within the subspace of initial excitation number $m_0$ becomes exponentially small and thus only a vanishingly small number of "successful" runs enter the measurement statistics. 

In order to maintain the dissipatively created off-diagonal order in the many-spin system, a crucial property of the excitation number measurement is its QND nature: the spin excitation number $m$ has to be determined in a way that allows one to only obtain information about the total number of excitations in the system, but no knowledge about the spatial positions of individual excitations along the array. This QND measurement can be realized by a global unitary map, which acts on the entire register of system spins and an ancillary qubit (see Fig.~5b of the main text). Such a unitary is constructed in a way that it maps the binary information whether the register of system spins is in the correct excitation number subspace with $m=m_0$ (or not) onto the logical state $\ket{0}$ ($\ket{1}$) of the ancillary qubit. It involves the projector $P_{m_0}^{(N)}$ onto the subspace of $m_0$ spin excitations in an array of $N$ spins, and reads
\begin{align}
\label{eq:U_n_map}
U_{m_0}^{(N)} & = \exp \left( - i\frac{\pi}{2} P_{m_0}^{(N)} \otimes \sigma_0^x \right) \\
& = P_{m_0}^{(N)} \otimes  (- i \sigma_0^x ) + (1- P_{m_0}^{(N)}) \otimes 1_0. \nonumber
\end{align}
This equation can be understood as follows: The state of the ancilla qubit, initially prepared in $\ket{1}$, is flipped by the $\sigma_0^x$ operation if the system spins are in a state with exactly $m=m_0$
excitations, whereas the ancilla qubit is left unchanged otherwise. The QND measurement is then completed by a measurement of the ancilla qubit in the computational basis, providing the desired information on whether $m=m_0$ or $m\neq m_0$, depending on whether the ancilla qubit is measured in $\ket{0}$ or $\ket{1}$.

\subsection{Construction of the projectors onto excitation number subspaces}
\label{subsec:P_construction}
The projector $P_{m_0}^{(N)}$ required for the unitary (\ref{eq:U_n_map}) can be constructed systematically and efficiently for any $m_0$ and $N$ as follows: One starts from the ansatz
\begin{equation}
P_{m_0}^{(N)} = \sum_{k=0}^N \alpha_k S_z^k 
\end{equation}
with $S_z = \sum_{i=1}^N \sigma_i^z$. This ansatz assures that the projector $P_{m_0}^{(N)}$ (i) is diagonal in the computational basis, (ii) symmetric under any permutation of spin indices, and (iii) does not involve higher powers with $k>N$ as such terms are already contained in previous terms with $k \leq N$ due to the property $(\sigma_i^z)^2 =1$. Since the computational basis states are eigenstates of $S_z$, 
\begin{align}
S_z \ket{0\ldots 0} & = N  \ket{0\ldots 0}, \nonumber \\
S_z  \ket{0\ldots 0, 1} &= (N-2)  \ket{0\ldots 0,1}, \ldots \nonumber \\
& \,\,\,\,\vdots \nonumber \\
S_z \ket{1\ldots 1} & = -N \ket{1 \ldots 1}, \nonumber
\end{align}
and the projector per definition fulfills
\begin{align}
P_{m_0}^{(N)} \ket{0 \ldots 0} & = 0, \nonumber \\
& \,\,\,\,\vdots \nonumber \\
P_{m_0}^{(N)}  \ket{\overbrace{0 \ldots 0}^{m_0}  \overbrace{1 \ldots 1}^{N-m_0}} & =  \ket{0 \ldots 0 1 \ldots 1}, \ldots \nonumber \\
& \,\,\,\,\vdots \nonumber \\
P_{m_0}^{(N)} \ket{1 \ldots 1} & = 0, \nonumber
\end{align}
its form is uniquely determined by the following coupled system of $N+1$ linear equations,
\begin{eqnarray}
\left( \begin{array}{ccccc} 1 & N & N^2 & \ldots & N^N \\
1 & (N-2) & (N-2)^2 & \ldots & (N-2)^N \\
\vdots & \vdots & \vdots & & \vdots \\
\vdots & \vdots & \vdots & & \vdots \\
1 & - N & (-N)^2 & \ldots & (-N)^N
\end{array} \right) \hspace{-2mm}
\left( \begin{array}{c}
 \alpha_0 \\ \alpha_1 \\ \vdots \\ \vdots \\ \alpha_N 
 \end{array} \right)
= \left( \begin{array}{c}
0 \\ \vdots \\ 1 \\ 0 \\ \vdots 
\end{array} \right)  \nonumber
\end{eqnarray} 
with the only non-zero entry in the $(N-m_0+1)$-th row. This matrix equation is readily solved, yielding for the experimentally relevant case of three system spins with one or two spin excitations the projectors
\begin{align}
\label{eq:supp_P1}
P_1^{(3)} & = \ket{011}\bra{011} +\ket{101}\bra{101} +\ket{110}\bra{110} \\
& = \frac{1}{16} (9 - 9 S_z - S_z^2 + S_z^3) \nonumber \\
& = \frac{1}{8} \left( 3- (\sigma_1^z + \sigma_2^z + \sigma_3^z) \right. \nonumber \\
& \qquad \left. - (\sigma_1^z \sigma_2^z + \sigma_1^z \sigma_3^z + \sigma_2^z \sigma_3^z) + 3 \sigma_1^z \sigma_2^z \sigma_3^z \right)\nonumber 
\end{align}
and
\begin{align}
\label{eq:supp_P2}
P_2^{(3)} & = \ket{001}\bra{001} +\ket{010}\bra{010} +\ket{100}\bra{100} \\
& = \frac{1}{16} (9 + 9 S_z - S_z^2 - S_z^3) \nonumber \\
& = \frac{1}{8} \left( 3 + (\sigma_1^z + \sigma_2^z + \sigma_3^z) \right. \nonumber \\
& \qquad \left. - (\sigma_1^z \sigma_2^z + \sigma_1^z \sigma_3^z + \sigma_2^z \sigma_3^z) - 3 \sigma_1^z \sigma_2^z \sigma_3^z \right), \nonumber
\end{align}
where $S_z = \sum_{i=1}^3 \sigma_i^z$. Note that the projectors are closely related and can be transformed into each other by the symmetry operation $S_z \rightarrow - S_z$, which interchanges the role of up- and down-spins, or occupied and empty sites in the hardcore boson model, respectively. 

\subsection{Experimental implementation of the QND measurement}
\label{subsec:QND_implementation}
As by the total spin excitation number $m$ a collective property of the entire spin system is measured, the unitary of Eq.~(\ref{eq:U_n_map}) truly is a many-qubit operation: Equations~(\ref{eq:supp_P1}) and (\ref{eq:supp_P2}) show that the projectors contains 1, 2 and 3-body spin interaction terms, such that the QND mapping of Eq.~(\ref{eq:U_n_map}) involves interaction terms with up to 4-body Pauli operators. A decomposition of the unitary for the QND measurement $U_{m_0=1}^{(3)}$ into experimentally available gates, as obtained using the numerical optimization algorithm, is shown in Table~\ref{tab:QND}. We note that since the QND measurement involves a \textit{global} unitary, a numerical optimization has to be done separately for any register size and any particular spin excitation number, and furthermore becomes inefficient for increasing register sizes. However, the unitary can be implemented efficiently without resorting to numerical optimization: For a system of $N$ spins, $U_{m_0}^{(N)}$ will generally be the product of unitaries corresponding to many-spin interactions, at most $(N+1)$-body Pauli operators. These unitaries can be decomposed into sequences of available gates following the recipes outlined in~\cite{mueller-njp-13-085007}. Although the implementation of  $U_{m_0}^{(N)}$ becomes experimentally demanding for increasing system sizes $N$ and in general requires more operations than numerically optimized circuits, the number of required operations for the QND mapping operation still grows polynomially with the number of system spins.
\begin{table}[h]
  \centering
  \begin{tabular}{|c|c||c|c|}
    \hline
    Number & Pulse & Number & Pulse\\
    \hline
1 & $R(0.5, -0.5)$  & 11 & $R(0.146, -0.895)$  \\  
2 & $R(0.5, 0.0)$  &12 & $MS(0.375, -1.054)$  \\ 
3 & $S_z(0.5, 3)$  & 13 & $S_z(0.364, 3)$  \\ 
4 & $MS(0.125, 0.0)$  & 14 & $MS(0.75, -1.054)$  \\ 
5 & $R(0.098, 1.0)$  & 15 & $R(1.0, 0.0)$  \\ 
6 & $S_z(1.636, 3)$  & 16 & $S_z(1.818, 3)$  \\ 
7 & $MS(0.25, 0.0)$  & 17 & $R(0.277, -1.054)$  \\ 
8 & $R(0.136, 0.5)$  & 18 & $S_z(0.152, 3)$  \\ 
9 & $S_z(0.75, 3)$  & 19 & $R(0.5, 0.895)$  \\ 
10 & $R(0.113, -1.054)$  & & \\
    \hline
  \end{tabular}
  \caption{Pulse sequence for the QND post-selective measurement of the spin excitation number in a system of 3+1 ions.}
  \label{tab:QND}
\end{table}

Due to the considerable complexity of the mapping operation for the QND measurement, with the optimized circuit involving in total 19 operations (see Table~\ref{tab:QND}), this method for error detection itself can only be implemented with a certain accuracy and requires a constant resource overhead, independently of the number of dynamical maps applied in the simulation. It can be seen from the data shown in Figs.~3a and b, and Fig.~4a of the main text, that for short sequences such as e.g.~only a single elementary dissipative map, where the system without error detection remains with high probability in the desired excitation number subspace, experimental imperfections in the QND measurement itself actually introduce even more errors on the state of the system (red data points) than in the case where it is not applied (blue data points). However, for longer sequences of dynamical maps, where the population loss out of the initial excitation number subspace becomes more and more significant, the application of the QND post-selective method becomes effective and enables a more accurate simulation of the system dynamics for longer times.

\section{Quantum error reduction scheme - Excitation number stabilization}
\label{sec:error_reduction} 

\subsection{General idea}

The second error reduction procedure goes beyond the error detection method described above: it does not only allow one to detect errors, which have changed the ideally conserved spin excitation number $m$ during the quantum simulation, but performs an active stabilization of the register of system spins in the wanted subspace of spin excitation number $m_0$. In the previous post-selective case the ancilla qubit carried the binary information whether the system is in the correct subspace or
not. Here, in contrast, it is necessary to distinguish between at least three cases: (i) The
excitation number $m$ is correct, and thus no error correction process is required; the excitation number is (ii) too
small or (iii) too large, as illustrated in figure 5b,c in
the main text. This information cannot be stored in a single ancilla qubit with two logical states anymore. 
Although it is in principle possible to use a higher-dimensional ancillary system, such as multiple ancillary qubits, to
store the required information, such an approach would require
significantly more complex detection and correction
algorithms. Here, we choose an alternative approach, which still allows one to perform the excitation number stabilization
with a single ancilla qubit if the stabilization process is implemented as a
two-step process: One stabilization step injects a single excitation into the register
if there are too few excitations present ($m<m_0$) and a second stabilization step removes an
excitation from the system if too many are present ($m>m_0$), as schematically shown in Fig.~5c of the main text. Similarly to repetitive quantum error correction, the state of the ancilla qubit has to be reset in
between the two steps. As the underlying construction of the excitation injection and the extraction step is very similar, we focus in the following on a detailed description of the protocol for the excitation extraction procedure.

Conceptually, the injection protocol consists of two main parts: First, similar to the QND measurement above, the binary information whether or not there are too many spin excitations in the system is coherently mapped onto the two logical states of an ancillary qubit. Second, conditional on the state of the ancilla qubit, which is acting as a quantum controller, a feedback procedure is applied to the system spins. This quantum feedback procedure extracts one of the (possibly multiple) superfluous spin excitations and stops once this has been achieved.
 
The first part is realized by a unitary mapping acting on the entire register of system spins and the ancilla qubit. However, in contrast to the QND post-selective method, the ancillary qubit is not measured after this mapping. The unitary for the mapping reads
\begin{align}
\label{eq:U_errcorr}
U_{m>m_0}^{(N)} & = \exp \left( - i\frac{\pi}{2}
\left( \sum_{j = m_0 +1 }^{N} P_j^{(N)} \right) \otimes \sigma_0^x \right)
\end{align}
and can be understood as follows: The state of the ancilla qubit prepared in $\ket{1}$ is flipped if and only if there are \textit{too many} spin excitations present in the system; here the system operator $\sum_{j=m_0+1}^N P_j^{(N)}$ denotes the projector onto the subspace of states containing strictly more excitations than the desired value $m_0$. It is the sum over all projectors $P_j^{(N)}$ onto subspaces with $j > m_0$ excitations in a system of $N$ spins, and each of these projectors can be readily constructed following the procedure described in Sec.~\ref{subsec:P_construction}.

After the mapping operation between the system spins and the ancilla, the second part of the protocol deals with the actual extraction of a spin excitation from the system. This step faces the difficulty that due to the QND character of the first mapping step, the state of the ancilla qubit only encodes information about whether too many excitations are present in the system, but not on which sites the excitations are located. In order to minimally disturb the state of the system spins it is desirable to devise an extraction scheme with the following properties: (i) An excitation is extracted in a minimally invasive way, i.e.~under an extraction of an excitation from a certain site, off-diagonal order among system spins of the rest of the chain is maintained as far as possible. (ii) Second, the scheme should effectively hold and not further alter the state of the system spins once an excitation has been successfully extracted. We have developed and implemented a scheme, which satisfies both criteria, and which exploits a combination of spectroscopic decoupling and optical pumping of the ancillary qubit. 

\begin{figure*}[h!]
\center
\includegraphics[angle=0,width=12cm]{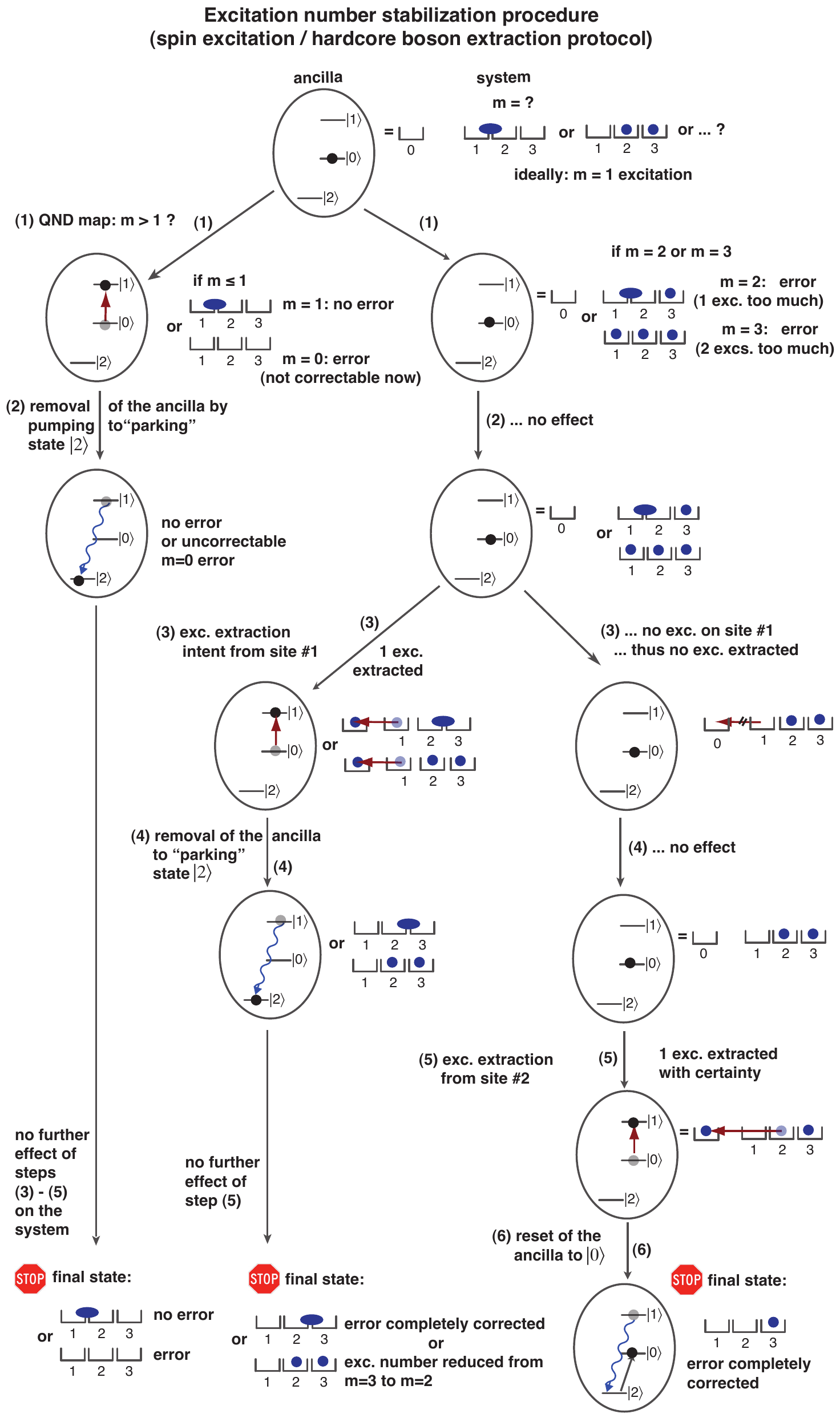}
\caption{Logical tree illustrating the complete excitation extraction procedure for a 3-spin system with ideally a single excitation present, and for various (unknown) initial states of the system spins. Blue circles correspond to spin excitations (hardcore bosons) localized on particular sites, whereas ellipses indicate situations where excitations (hardcore bosons) can be delocalized over several lattice sites. The error reduction takes place in six steps: By a QND map (1) the information whether or not too many excitations are present in the system, is mapped onto an ancilla qubit. (2) In case of the ideal excitation number or too few excitations, the ancilla is removed from the computational space and the protocol effectively halts. In the opposite case, a swap operation (3) is applied to try to extract an excitation from system site \# 1. If successful, after the subsequent removal of the ancilla (4) the protocol halts. Otherwise, (5) an extraction by a swap of the ancilla with system site \# 2 extracts an excitation. (6) Finally, the ancilla is reset to its initial state $\ket{0}$ for subsequent error reduction steps. See also text for a more detailed explanation.}
\label{fig:QEC_boson_extraction}
\end{figure*}

\subsection{Steps of the excitation extraction procedure} 
\label{sec:steps_extraction}
Let us now describe in more detail the steps of the excitation extraction procedure, which is
illustrated in Fig.~\ref{fig:QEC_boson_extraction} for a system of three spins. In step~(1) the
mapping introduced above is applied, so that the information whether a
correctable error (i.e. too many spin excitations in the system) is present (or not), is encoded in the ancilla state $\ket{0}$ ($\ket{1}$). If no correctable error is detected ($m \leq m_0$) then the subsequent
operations should not affect the system qubits. This can be realized
by removing the ancilla qubit incoherently from the computational subspace
(step~(2)) as will be discussed below. 
In step~(3) an extraction attempt on the first site of the spin register is performed by swapping the ancilla, which
is now in the state $\ket{0}$ corresponding to an empty site (no spin excitation), with the first
qubit of the system. In terms of spins this operation corresponds to a swap ("flip-flop") process, under which $\ket{0}_0 \ket{1}_1 \rightarrow \ket{1}_0 \ket{0}_1$, i.e.~the spin excitation is coherently exchanged between the ancilla qubit and the first system spin. In the language of hardcore bosons, the process can be interpreted as a tunneling of a boson from the site \#1 onto the previously empty ancilla site (see step (3) in Fig.~\ref{fig:QEC_boson_extraction}). If after this process the ancilla qubit is in the state corresponding to an occupied site (spin excitation present), a single excitation has been successfully extracted from the system and
the process should halt. As will be shown below, this can be
performed by removing the population of the ancilla qubit from the computational
subspace (step (4)). However, if the ancilla is still in the state $\ket{0}$ corresponding to an empty 
site after the swap operation, then the first system qubit was in the state corresponding to an unoccupied site and thus
no spin excitation (or hardcore boson) could extracted, $\ket{0}_0 \ket{0}_1 \rightarrow \ket{0}_0 \ket{0}_1$. Thus, this
procedure is repeated (step (5)) until the ancilla qubit is found in the state $\ket{1}$ corresponding to an 
occupied site. In the case of three sites at most two extraction rounds
are required (steps~(3), (4), and (5)), before the ancilla is reset to $\ket{0}$ (step (6)) for subsequent rounds of error reduction. 

\textit{Open vs. closed-loop error reduction scheme} -- The described extraction (and injection) protocols are realized in an \textit{open-loop} fashion, i.e.~the ancilla remains unobserved during the whole procedure. This comes at the cost that all pulses (for steps (1) - (6) in the above example) have to be physically applied in every run of the error reduction procedure, even if the ancilla has been already be removed from the computational subspace and in principle no further operations would be required. However, the described protocol could easily be modified into a \textit{closed-loop} control scenario: In this case one would perform individual measurements of the ancilla qubit after the QND map (1) as well as after the coherent swap operations (steps (3) and (5)). The outcomes of these ancilla measurements would then yield the classical information on whether no error correction is required after step (1) or an excitation has already been extracted successfully in steps (3) or (5). This information could then be used for classical feedback on the quantum system in the sense that one can externally decide whether further steps are required in the error reduction protocol, or whether one can stop the current run of the error reduction protocol and no further operations have to be applied. An advantage of this modified scheme is that dissipative removal operations of the ancilla qubit out of the computational subspace are not required. Experimentally, the price to pay in this modified scheme would be a number of fluorescence measurements, which are slow compared to the application of coherent gates, and the requirement to recool the relevant external degrees of freedom of the ion chain after each measurement as recently demonstrated in~\cite{Schindler12}. 

\subsection{Experimental implementation}
In the experiment, the error reduction protocol for a system of three spins, using 3+1 ions was implemented. We considered the case of having ideally $m_0=1$ excitation present on the three sites. For the extraction step, the four-qubit unitary $U_{m>1}^{(3)}$ (cf.~Eq.~(\ref{eq:U_errcorr})) to detect whether too many excitations (i.e.~$m=2,3$) are present in the system or not, has been obtained by numerical optimization. The resulting circuit consisting of 15 gates is listed in Table~\ref{tab:removal}. On the other hand, for the injection part, the unitary   \begin{align}
\label{eq:U_err_corr_injection}
U_{m<m_0}^{(N)} & = \exp \left( - i\frac{\pi}{2} \left(
\sum_{j=0}^{m_0 -1} P_j^{(N)} \right) \otimes \sigma_0^x \right)
\end{align}
which interrogates the system whether too few excitations are present, is required. For the case $m_0=1$ and $N=3$, the unitary $U_{m<1}^{(3)}$ realizes a spin flip of the ancilla qubit only if the three system spins are in the state $|000\rangle$. This operation is equivalent to a triple controlled-NOT operation with the three system spins playing the role of the control qubits and the ancilla the target qubit. As we could not obtain a satisfying circuit decomposition for this unitary by means of the usual numerical optimization algorithm, we did not try do directly optimize $U_{m<1}^{(3)}$, but allowed the unitary to add arbitrary phases to states lying outside the desired excitation number subspace $m_0=1$. Under these relaxed conditions, the numerical optimizer delivered a circuit decomposition of 19 operations as shown in Table~\ref{tab:injection}.

\begin{table}[h]
  \centering
  \begin{tabular}{|c|c||c|c|}
    \hline
    Number & Pulse & Number & Pulse \\
    \hline
1 & $R(0.25, 0.5)$  & 9 & $S_z(1.0, 3)$  \\
2 & $S_z(1.0, 3)$  & 10 & $R(0.25, 0.5)$  \\ 
3 & $R(0.25, 0.5)$  & 11 & $MS(0.5, 0.5)$  \\ 
4 & $MS(0.25, 0.5)$  &12 & $R(1.5, 0.5)$  \\ 
5 & $R(1.75, 0.0)$  & 13 & $R(0.125, 0.5)$  \\ 
6 & $S_z(1.0, 3)$  & 14 & $S_z(1.0, 3)$  \\ 
7 & $R(0.25, 0.0)$  &15 & $R(1.875, 0.5)$  \\ 
8 & $R(1.75, 0.5)$  &  & \\
    \hline
  \end{tabular}
  \caption{Pulse sequence to implement the QND unitary $U_{m>1}^{(3)}$ (cf.~Eq.~(\ref{eq:U_errcorr})) as part of the excitation extraction protocol. It maps the information whether there are more than $m=1$ spin excitations present in a system of three spins, onto the ancilla qubit.}
  \label{tab:removal}
\end{table}

\begin{table}
  \centering
  \begin{tabular}{|c|c||c|c|}
    \hline
    Number & Pulse &Number & Pulse  \\
    \hline
1 & $R(1.75, 0.5)$  & 12 & $R(0.375, 0.0)$  \\
2 & $S_z(1.0, 3)$  & 13 & $S_z(1.0, 3)$  \\ 
3 & $R(1.75, 0.5)$  & 14 & $R(1.625, 0.0)$  \\ 
4 & $MS(0.25, 0.0)$  & 15 & $S_z(0.5, 3)$  \\ 
5 & $R(1.875, 0.0)$  & 16 & $MS(0.25, 0.0)$  \\ 
6 & $S_z(1.0, 3)$  & 17 & $R(0.125, 0.0)$  \\ 
7 & $R(0.125, 0.0)$  & 18 & $S_z(1.0, 3)$  \\ 
8 & $R(0.25, 0.5)$  & 19 & $R(1.875, 0.0)$  \\ 
9 & $S_z(1.0, 3)$  &  20 & $R(0.125, 0.5)$  \\  
10 & $R(1.75, 0.5)$  & 21 & $S_z(1.0, 3)$  \\ 
11 & $MS(0.25, 0.0)$  & 22 & $R(1.875, 0.5)$  \\ 
    \hline
  \end{tabular}
   \caption{Pulse sequence to implement the QND unitary $U_{m<1}^{(3)}$ (cf.~Eq.~(\ref{eq:U_err_corr_injection})) as part of the excitation injection protocol. It maps the information whether there are $m=0$ instead of ideally $m=1$ spin excitations present in a system of three spins, onto the ancilla qubit.}
  \label{tab:injection}
\end{table}

\begin{figure}[h!]
\center
\includegraphics[angle=0,width=7cm]{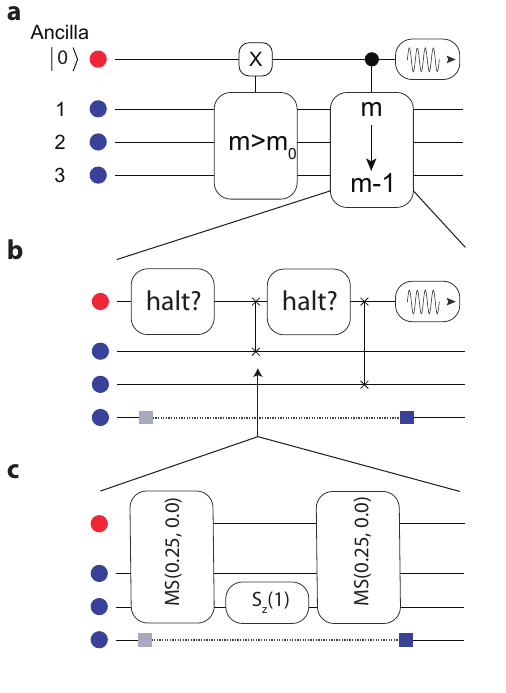}
\caption{{\textbf{a,}} Schematic view of the error detection and
  excitation removal process (cf.~Fig.~5c of the main text). {\textbf{b,}} The excitation removal process on 3
  system qubits can be performed by swap operations and effective
  halting conditions realized by a dissipative decoupling of the ancilla
  qubit. {\textbf{c,}} Swap operations (\ref{eq:flip_flop_operation}) between the ancilla qubit and one system spin are implemented by two
  effective two-qubit MS operations, which are build up from 4 MS operations acting on three ions, 
  interspersed with refocusing operations.}
\label{fig:schem_removal}
\end{figure}

\begin{figure}[h!]
\center
\includegraphics[angle=0,width=6.8cm]{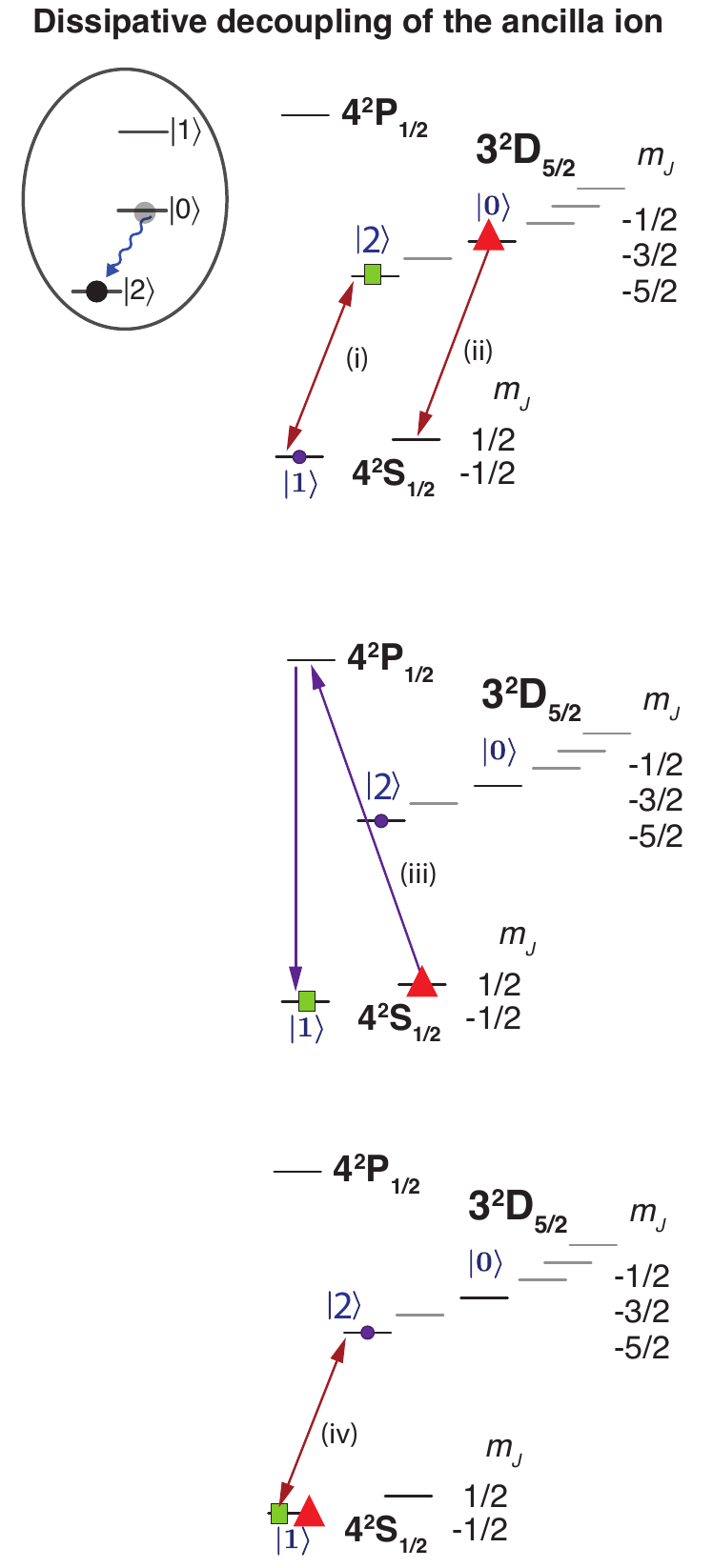}
\caption{The dissipative decoupling process of the ancilla qubit used in the
  injection/removal processes. The goal is to uni-directionally remove population in the computational state $\ket{0}$ (red triangular marker) and add it to electronic population that is possibly already present in the "parking" state $\ket{2}$ (green rectangular marker). Population in the computational state $|1\rangle$ (blue circle) should be left unchanged. The sequence: (i) First, the population in states $\ket{1}$ and $\ket{2}$ is coherently swapped by a $\pi$-operation. (ii) Population in $|0\rangle$ is coherently transferred to the state
  $4S_{1/2}(m_j=+1/2)$. (iii) Optical pumping from this state towards the
  $|1\rangle$ state. In this step, the populations that were at the beginning of the sequence present in $\ket{2}$ and $\ket{0}$ are added up and temporarily accumulate in state $\ket{1}$. (iv) Finally the populations in states $\ket{1}$ and $\ket{2}$ are again swapped coherently.}
\label{fig:qec_decouple}
\end{figure}

\textit{Implementation of the actual excitation extraction} -- We will now discuss the physical mechanism underlying the implementation of the removal of a spin excitation (or hardcore boson), as described above and illustrated in
Fig.~\ref{fig:schem_removal}. This excitation removal step takes place after the QND mapping (1) and a decoupling operation on the ancilla qubit (2) described below. As mentioned above, the extraction step (3) relies on a swap
operation exchanging the excitation of the ancilla qubit with one of the system spins, say \#1. This swap operation is realized by applying a $\pi$-pulse, according to the "flip-flop" Hamiltonian 
\begin{equation}
H_{01} = \sigma_0^+ \sigma_1^- + \sigma_0^- \sigma_1^+ = \frac{1}{2} (\sigma_0^x \sigma_1^x + \sigma_0^y \sigma_1^y). \nonumber
\end{equation}
The resulting unitary
\begin{align}
\label{eq:flip_flop_operation}
U_{01} = & \exp \left( -i \frac{\pi}{2} H_{01} \right) \\
= & \exp \left( - i \frac{\pi}{4} \sigma_0^x \sigma_1^x  \right) \hspace{-0.5mm}\exp \left( - i \frac{\pi}{4} \sigma_0^y \sigma_1^y  \right) \nonumber
\end{align}
corresponds to the application of two fully-entangling $x$- and $y$-type MS gates (see Eq.~(\ref{eq:MS_gate})) applied to the ancilla ion and the first system ion. This two-ion MS gate can be realized by the global bi-chromatic light field, if all ions but the ancilla qubit and the system qubit \#1 are spectroscopically decoupled. However, since step (3) involves the ancilla qubit and system ion \#1, and step (5) the ancilla qubit and system ion \#2, it is from an experimental point of view more convenient to spectroscopically decouple only ion \#3, but keep \textit{both} ions \#1 and \#2 in the "active" qubit states during the extraction procedure. Doing so, each of the $x-$ and $y$-type MS-gates required for the operation (\ref{eq:flip_flop_operation}) acting on the ancilla (index \#0) and system ion \#1 can be realized by two MS gates acting on the three ions, interspersed by a $\pi$-pulse $S_Z(1,2)$ applied to the second system ion, as shown in Fig.~\ref{fig:schem_removal}. The application of the "refocusing" pulse in between the MS gates leads to an effective decoupling of ion \#2 that is not supposed to participate in the unitary (\ref{eq:flip_flop_operation}) \cite{mueller-njp-13-085007}. This technique is also applied to realize the swap operation required in step (5), which acts on the ancilla qubit and the second system. Using this decoupling technique based on refocusing pulses, no additional
spectroscopic decoupling operations are required between the two swap
operations corresponding to steps (3) and (5). The experimentally employed sequence of 6 gates is listed in Table~\ref{tab:SWAP}.

\begin{table}
  \centering
  \begin{tabular}{|c|c|}
    \hline
    Number & Pulse \\
    \hline
1 & $MS(0.25, 0)$   \\ 
2 & $S_z(1.0 , j)$  \\ 
3 & $MS(0.25, 0)$ \\ 
4 & $MS(0.25, 0.5)$   \\ 
5 & $S_z(1.0 , j)$  \\ 
6 & $MS(0.25, 0.5)$ \\ 
    \hline
  \end{tabular}
  \caption{Pulse sequence for a swap operation used for excitation injection or removal, see Fig.~\ref{fig:schem_removal}c. Ion $j$ is the spectator ion, which is not spectroscopically decoupled, but is not intended to participate in the swap operation.}
  \label{tab:SWAP}
\end{table}

In Sec.~\ref{sec:steps_extraction} we have outlined that the ancilla will be removed from the computational subspace in step (2) if no error is to be corrected in the current error reduction round. In this case, if the ancilla has been removed from the qubit subspace, the two-qubit MS interactions appearing in (\ref{eq:flip_flop_operation}) act only on a single qubit. However, in this "pathological" case where the bi-chromatic laser fields used for the generation of the effective spin-spin interactions of the entangling MS gate of Eq.(\ref{eq:MS_gate}), are applied to a single ion only, these realize up to negligible corrections the identity operation on this single ion \cite{molmer-prl-82-1835,roos-njp-10-013002}. As a consequence, in this case -- despite the fact that the MS gate laser pulses of steps (3) and (5) are physically applied to the ions -- they do not alter the state of the systems spins, as desired. 

\textit{Dissipative decoupling of the ancillary qubit} -- As explained above, the ancilla qubit shall be removed from the computational subspace if either (i) no correctable error is detected in the current round of error reduction, or (ii) a superfluous spin excitation has been successfully extracted from the spin system. In these cases, the removal of the ancilla from the qubit subspace guarantees that the error reduction protocol effectively halts (see the logical tree shown in Fig.~\ref{fig:QEC_boson_extraction}) and the state of the system spins is no further modified, even though after the removal of the ancilla MS gate laser pulses are applied to the ion string. In steps (2) and (4) of the spin extraction protocol, the ancilla qubit is removed from the computational state $\ket{1}$ into the "parking" state $\ket{2}$, which we encode in the $3D_{5/2}(m_j=-5/2)$ electronic state (see Fig.~\ref{fig:qec_decouple}). This removal into $\ket{2}$ takes place if the ancilla ion resides at these instances in $\ket{1}$, if it is in $\ket{0}$ its state remains unaffected.

For this removal it is not possible to use the \textit{coherent} spectroscopic decoupling technique as used for the implementation of the elementary Kraus maps. The reason is that this would lead to errors in the protocol: for instance, imagine no correctable error is present in the system and thus the ancilla is brought to the decoupling state $\ket{2}$ in step (2). If then another ancilla removal were performed coherently (step (4)), the ancilla would be transferred back from $\ket{2}$ to the computational state $\ket{1}$. This is clearly unwanted as in this case with the ancilla returned to $\ket{1}$, the subsequent swap operation (5) would be performed by mistake. This unwanted behavior can be avoided if the transfer of the ancilla from $\ket{1}$ to $\ket{2}$ is realized \textit{dissipatively}, by an optical pumping process, which bares similarities with the incoherent reset of the ancilla qubit for the elementary dissipative maps. Such uni-directional, incoherent pumping process from $\ket{1}$ to $\ket{2}$ guarantees that once the ancilla has reached the "parking" state $\ket{2}$, it will in subsequent steps never return to the computational subspace. This optical pumping process for a removal process of the ancilla qubit from one of the computational basis states into $\ket{2}$ is illustrated and described in more detail in Fig.~\ref{fig:qec_decouple}. 

In the spin excitation protocol the ancilla is removed from $\ket{1}$ (corresponding to an occupied site) to $\ket{2}$, whereas in the spin injection protocol it has to removed from $\ket{0}$ (empty site) to $\ket{2}$. For both scenarios, the pulse sequence outlined in Fig.~\ref{fig:qec_decouple} can be employed; a $\pi$-pulse resonant with the qubit transition of the ancilla, applied before and after the dissipative decoupling sequence, exchanges the roles of the two computational states $\ket{0} \leftrightarrow \ket{1}$ and thereby allows one to switch between the spin extraction and injection scenario. 

\subsection{Experimental results for the stabilization}
\label{sec:exp_stabilization}

The active excitation number stabilization procedure can be best tested when applying it
to a state that has a considerable amount of the population outside
the subspace with the correct excitation number $m_0$. We applied Hadamard operations
on the three system qubits initially in $\ket{000}$ to prepare the initial state
$\ket{\psi_0} = 1/\sqrt{8}(|0\rangle+|1\rangle)^{\otimes 3}$, which is an equal-weight superposition of all eight three-qubit computational basis states, each occurring with probability 1/8. The measured and ideal density matrix of this initial state
is shown in Fig.~\ref{fig:qec_rho}a. For $m_0=1$ the states
$|001\rangle$, $|010\rangle$, $|100\rangle$ span the subspace with the desired spin excitation number. Thus, the initial fraction of population in this subspace is 3/8 as shown in Fig.~5d of the main text. We then performed the excitation extraction protocol according to the protocol outlined above and summarized in Fig.~\ref{fig:QEC_boson_extraction}. Ideally this protocol extracts one excitation from the components of the initial state, which contain two or three spin excitations, $m=2 \rightarrow m=1$ and $m=3 \rightarrow m=2$, thereby pumping the population corresponding to these states into the subspaces with one excitation less. It is a crucial property of the error reduction protocol that the coherences within the subspace of the "correct" excitation number $m=1$ are ideally preserved, as the dynamics within the desired simulation subspace should be affected as little as possible. The component $\ket{000}$ of the initial state corresponds to a state with zero, i.e.~too few spin excitations, and thus to an error which is not corrected by the spin excitation procedure. 

In Fig.~5d in the main text, the ideal and the measured populations in all four excitation number 
subspaces at the end of the extraction protocol are shown. From this information one can deduce that the protocol within experimental accuracy realizes the pumping of populations between the different excitation number subspaces as expected. To infer whether the coherences are preserved, we measured the three-qubit density matrices after the first excitation extraction attempt from the first site (step (3) in Fig.~\ref{fig:QEC_boson_extraction}) and after the second extraction attempt from the second site (step (5)). The measured and the ideal density matrices are shown in Fig.~\ref{fig:qec_rho}b and c, where the relevant coherences within the $m_0=1$ subspace are highlighted in red color. From the comparison of measured and ideal density matrices it can be
seen that the coherences in the $m_0$ excitation subspace are well-preserved.

The complete error correction protocol consists of the excitation extraction and injection procedures. For the same initial state $\ket{\psi_0}$, we implemented the injection procedure, which ideally only acts on the $\ket{000}$ state, and pumps the population from the $m=0$ into the $m=1$ subspace. Again, the coherences within the $m=1$ subspace should be preserved under this procedure, yielding as a result of the injection protocol the ideal density matrix shown in the right part of Fig.~\ref{fig:qec_rho}d. Comparison with the measured density matrix shows that most of the population is pumped out of the $m=0$ subspace, and that the initially present coherences within the $m=1$ subspace are reasonably well conserved. 

\begin{figure}[h]
  \centering
  \includegraphics[width=1\columnwidth]{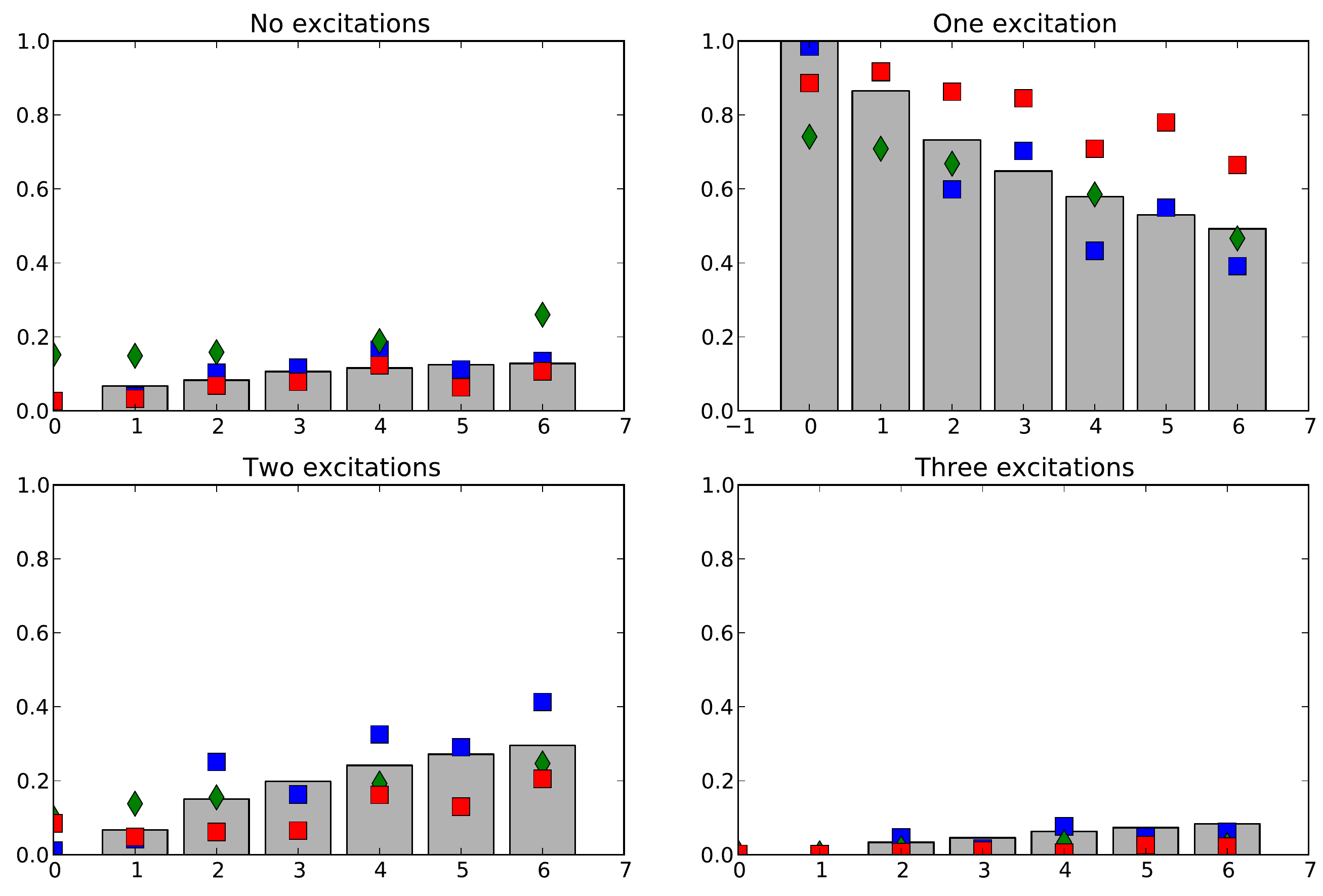}    
  \caption{Experimental study of the error reduction protocol, incorporated into the actual quantum simulation of composite dynamical maps with 3+1 ions, with an initial and ideal excitation number of $m=1$. The four plots show the fraction of the population in the subspaces with zero, one, two and three spin excitations, as a function of the number of elementary dissipative dynamical maps applied. For comparison, blue data points corresponds to the population under dissipative maps, without error detection or reduction technique applied. Red data points correspond to the case where after the final elementary dissipative map the QND post-selective measurement was applied. Green data points corresponds to the case where the extraction protocol has been applied at the end of the simulation. Bars correspond to the theory, where depolarizing noise in the elementary dissipative maps is taken into account (cf.~Appendix~\ref{sec:modeling_the_process}).}
  \label{fig:qec_insequence}
\end{figure}

\begin{figure*}[p]
  \centering
\includegraphics[width=\textwidth]{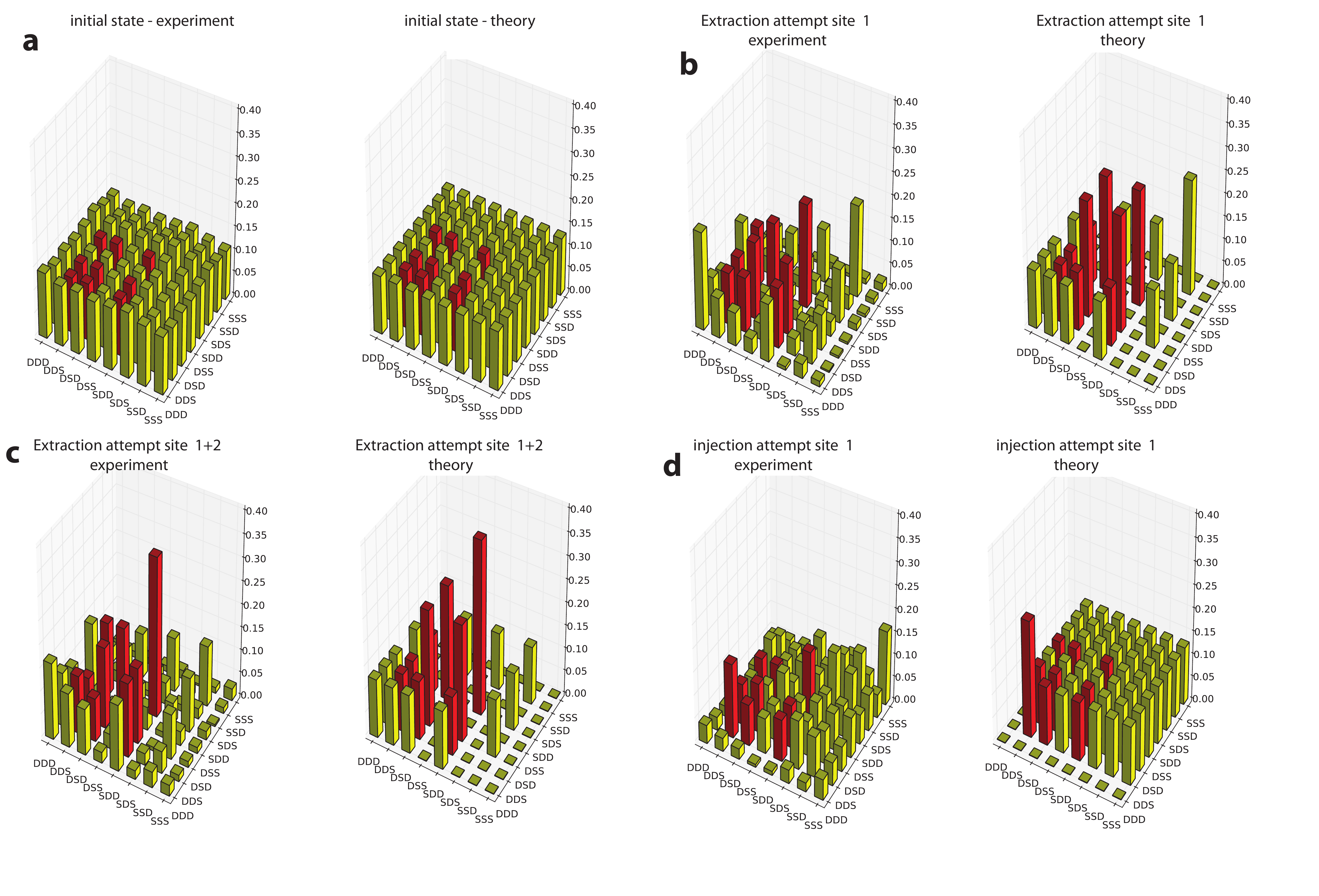}
  \caption{Reconstructed (left) and ideal (right) density matrices of the removal and injection
    process in a 3+1 ion experiment. Populations and coherences within the
    desired excitation number subspace with $m=1$ are high-lighted by red bars. {\textbf{a,}} The initial
    state is an equal-weight superposition of all eight computational basis states. {\textbf{b,}} The state after the system after a spin excitation removal attempt from the first site. {\textbf{c,}} The state of the system
    after the second swap operation to remove an excitation from the second site. After both steps, the coherences shown in red are well-preserved. {\textbf{d,}} The state of the
    system after a spin excitation injection attempt, starting again in the equal-weight superposition state shown in a. The population from the $\ket{DDD}$ = $|000\rangle$ state is depleted, and coherences within the $m=1$ subspace are reasonably well preserved.}
  \label{fig:qec_rho}
\end{figure*}

The ultimate goal of any error reduction protocol is certainly to
increase the performance of a complex algorithm. As a step in this direction, we integrated the
excitation removal protocol into the simulation sequence for dynamics according to composite
dissipative maps with 3+1 ions. In Fig.~\ref{fig:qec_insequence} we compare the
probabilities for all excitation numbers $m$ when (i) no error
reduction technique is used (blue data points), (ii) with the post-selective QND
measurement applied (red), and (iii) with the excitation removal procedure included in the simulation (green). It can
be seen that the removal procedure has a higher overhead due to its
considerable complexity. Nevertheless, as an indication of its
usefulness, a slower decay of the probability of finding the population in the desired subspace for $m=1$ can be observed, for the case in which the stabilization procedure is applied, compared to the case without any error correction. This indicates that the stabilization procedure indeed works qualitatively as intended, when it is incorporated into the actual simulation.  

  \begin{figure}[!h]
    \centering
    \includegraphics[width=\columnwidth]{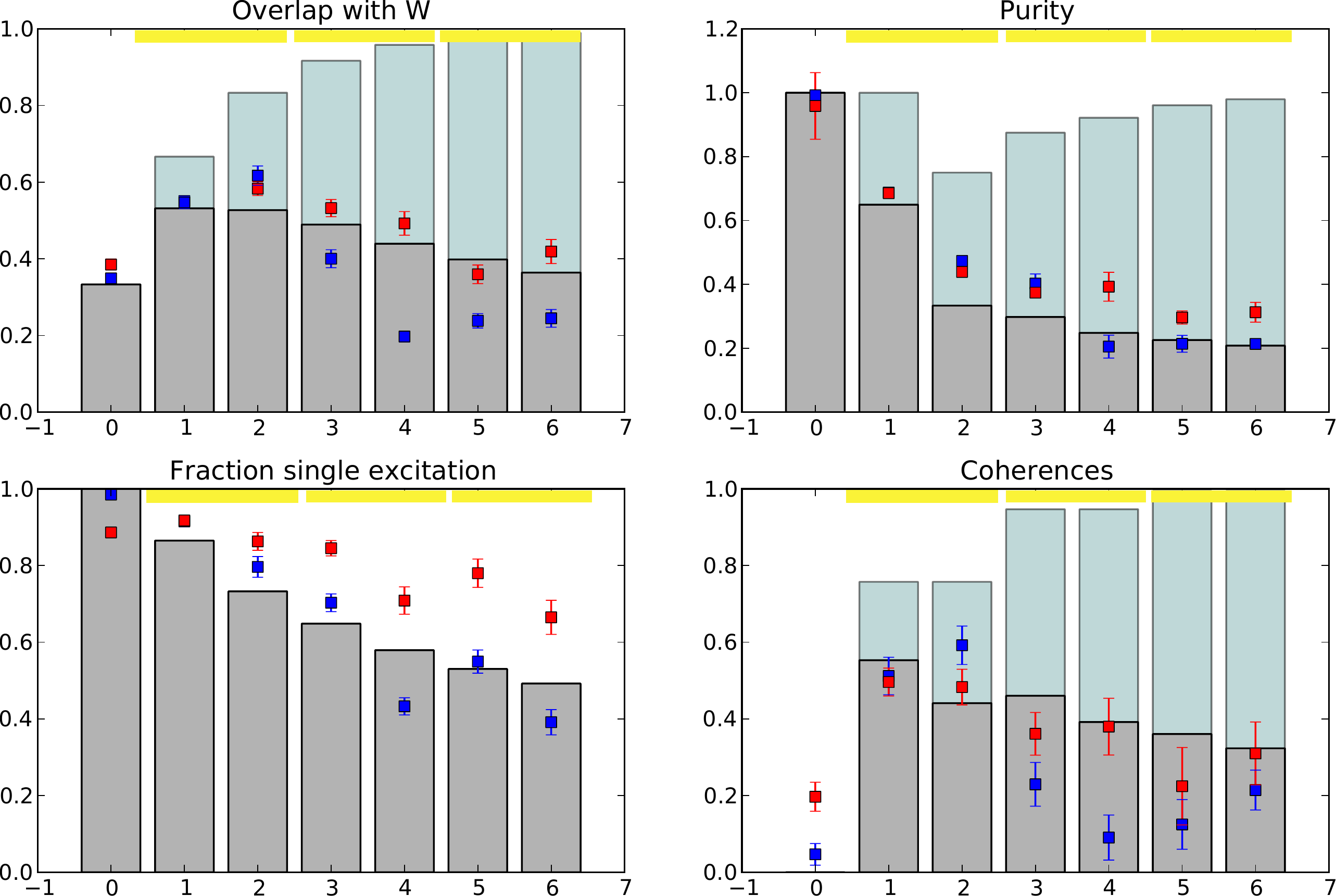}    
    \caption{ {\textbf{Experimental results of dissipatively induced
          delocalization through composite dynamical maps with 3+1
          ions.}} The results from an ideal model are shown in
      light-blue bars whereas those from a model including
      depolarization noise are indicated by dark-grey bars. Blue
      rectangles indicate the experimentally observed dynamics without
      any correction scheme whereas red rectangles include a
      post-selective error detection scheme (error bars,
      1$\sigma$). Overlap fidelity, purity, population in the $m=2$
      subspace, and off-diagonal order in a 4-spin quantum simulation
      with 4+1 ions, studying purely dissipative dynamics that induces
      pumping towards the Dicke state $\ket{D(2,4)}$ as shown in
      figure 3 in the main text.}
    \label{fig:all_pumping}
  \end{figure}
  \begin{figure}[!h]
    \centering
    \includegraphics[width=\columnwidth]{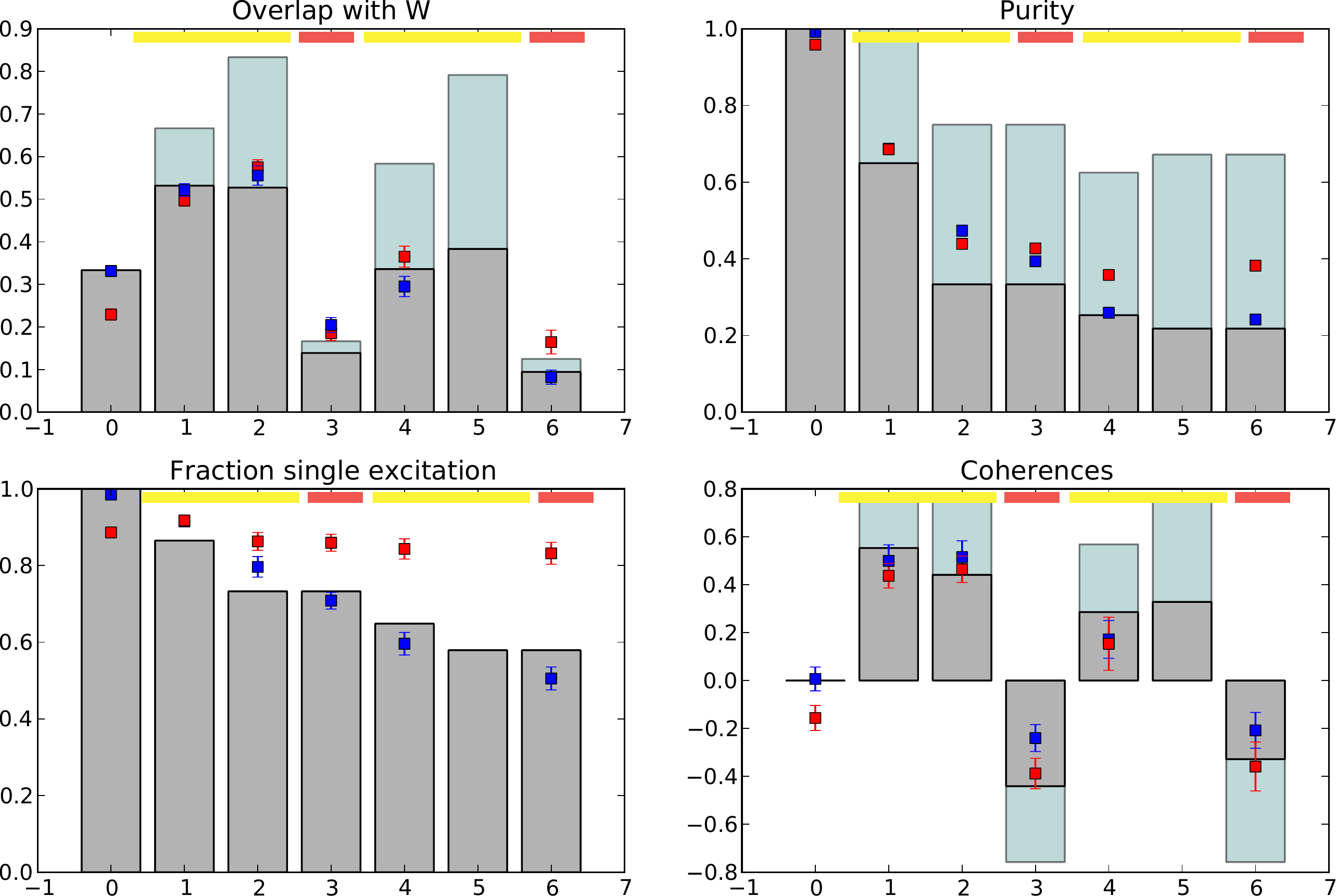}
    \caption{ {\textbf{Experimental results for competing dissipative
          and coherent dynamics with 3+1 ions}} The results from an
      ideal model are shown in light-blue bars whereas those from a
      model including depolarization noise are indicated by dark-grey
      bars. Blue rectangles indicate the experimentally observed
      dynamics without any correction scheme whereas red rectangles
      include a post-selective error detection scheme (error bars,
      1$\sigma$).  Overlap fidelity, purity, population in the $m=2$
      subspace, and off-diagonal order in a 3-spin quantum simulation
      with 3+1 ions. The dynamics corresponds to dissipative maps and
      coherent Hamiltonian competition as shown in figure 4 in the
      main text.}
    \label{fig:all_comp}
  \end{figure}
  \begin{figure}[!h]
    \centering
    \includegraphics[width=\columnwidth]{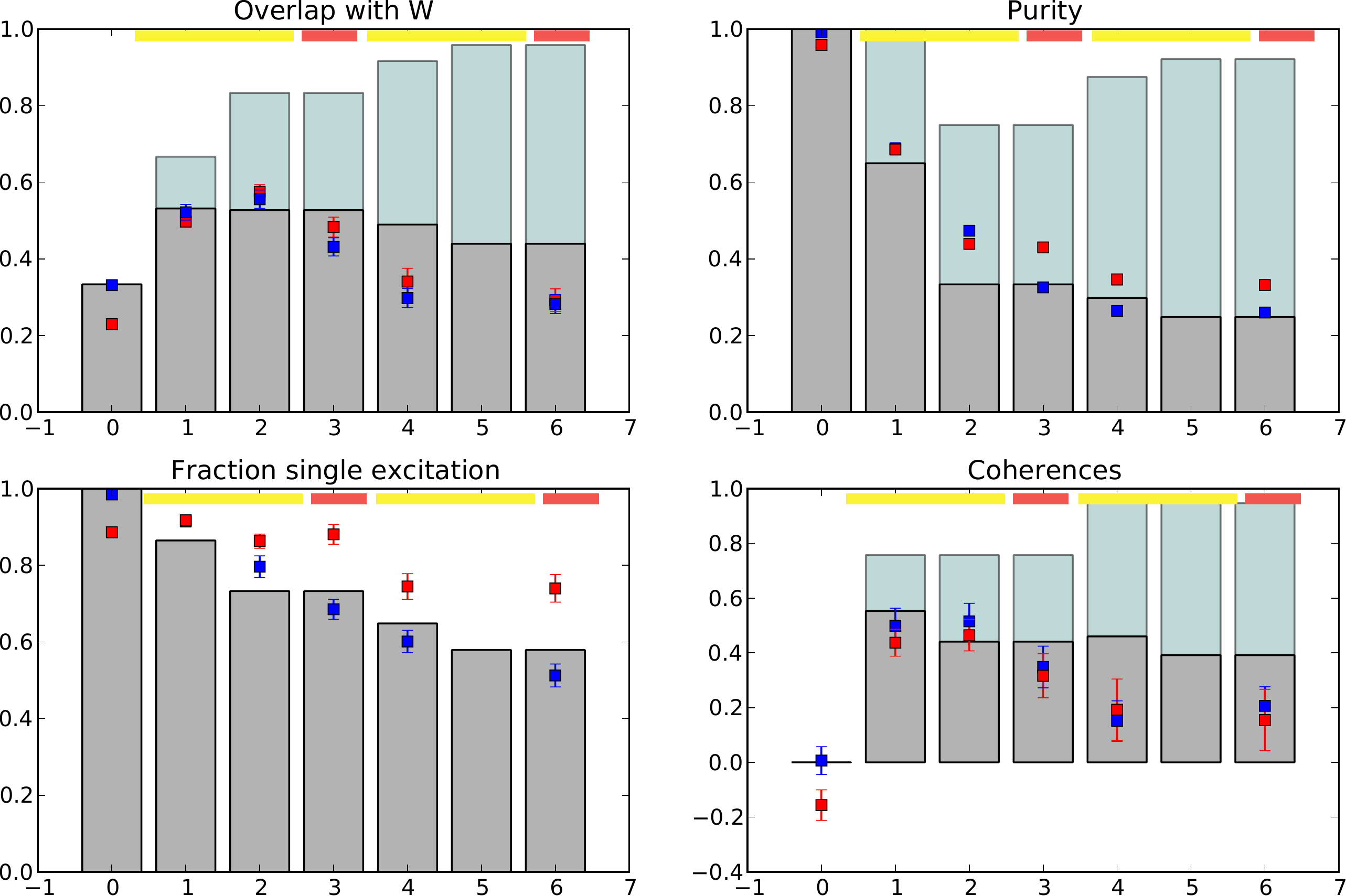}    
    \caption{{\textbf{Experimental results for competing dissipative
          and coherent dynamics with 3+1 ions with a single
          excitation.}} The results from an ideal model are shown in
      light-blue bars whereas those from a model including
      depolarization noise are indicated by dark-grey bars. Blue
      rectangles indicate the experimentally observed dynamics without
      any correction scheme whereas red rectangles include a
      post-selective error detection scheme (error bars,
      1$\sigma$). Data from simulated dynamics with Hamiltonian
      competition for 3 spins (3+1 ions), but only a single spin
      excitation present. As physically expected, the data confirms
      that in this case the Hamiltonian dynamical map has no effect.}
    \label{fig:single_boson}
  \end{figure}
  \begin{figure}[!h]
    \centering
    \includegraphics[width=\columnwidth]{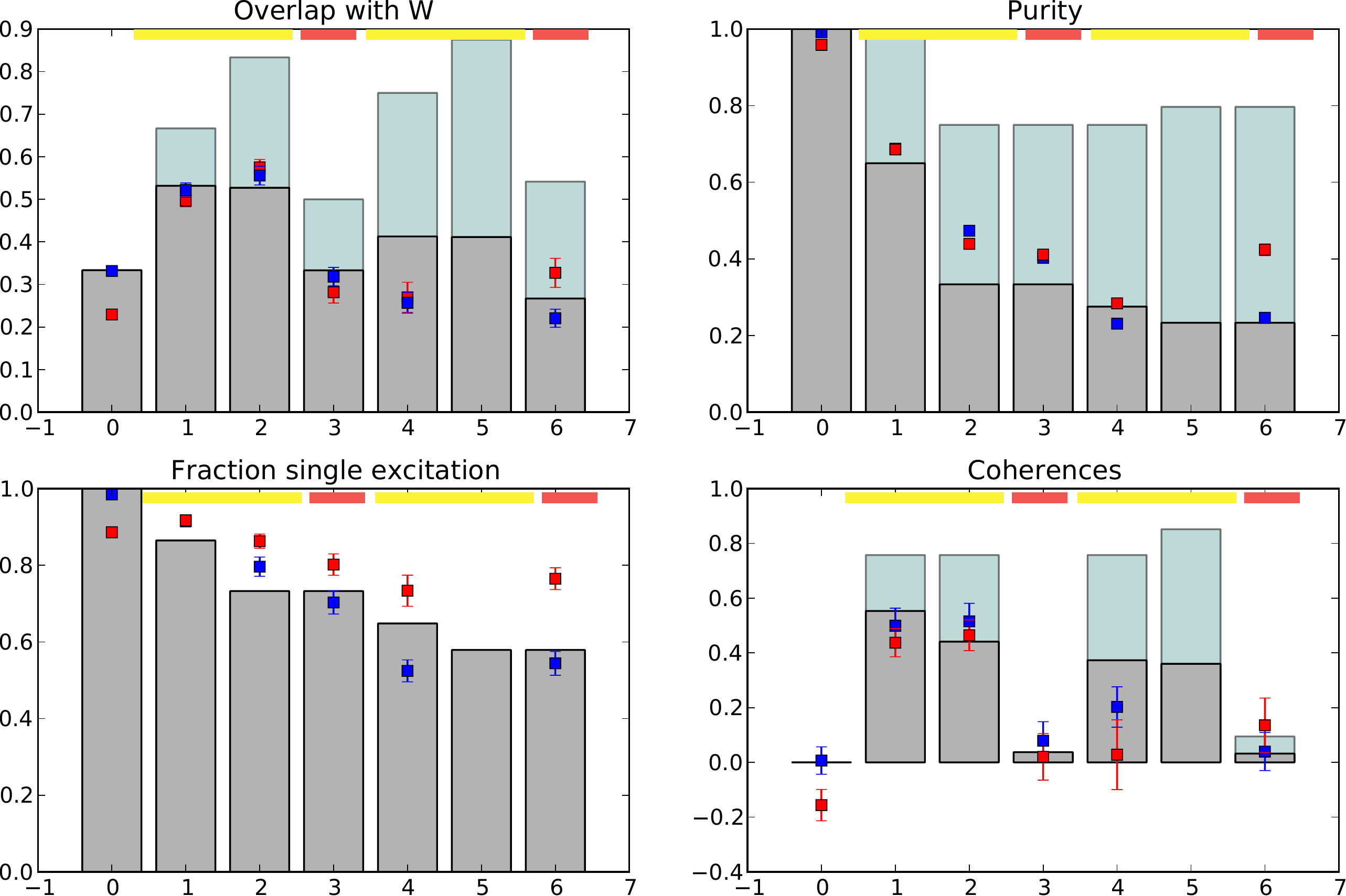}    
    \caption{{\textbf{Experimental results for competing dissipative
          and coherent dynamics with 3+1 ions with weak interaction}}
      The results from an ideal model are shown in light-blue bars
      whereas those from a model including depolarization noise are
      indicated by dark-grey bars. Blue rectangles indicate the
      experimentally observed dynamics without any correction scheme
      whereas red rectangles include a post-selective error detection
      scheme (error bars, 1$\sigma$). Data from simulated dynamics
      with weaker competing Hamiltonian dynamics ($\phi= \pi/4$) in a
      3-spin system with two spin excitations present in the system
      (3+1 ions).}
    \label{fig:weak_comp}
  \end{figure}
\section{Additional experiments and data analysis for competing dissipative and coherent dynamics}
\label{sec:additionalexp} 
In Figs.~3 and 4 of the main text, the experimental results
for systems of 3+1 and 4+1 ions are depicted. The attentive reader
noticed that in Fig. 4a in the main text the data point for 5
maps is missing. This is due to the fact that in view of the length of the algorithm, the memory of the current experimental control
system is not sufficient to generate the required pulse sequence for 4 elementary
dissipative maps and one composite Hamiltonian dynamical map. It is however possible
to generate the sequence for two elementary maps and one Hamiltonian dynamical map, and to repeat this sequence twice (data point for 6 maps).

Here, we extend the experimental analysis that was omitted from the
main text by showing different measures for
the already presented data and also additional datasets:
In Figs.~\ref{fig:all_comp} and~\ref{fig:all_pumping} we add the
measures purity and off-diagonal order for the data already presented
in Figs.~3 and 4 of the main text. The purity $\mathrm{Tr}\rho^2$ is a measure for how close the
measured state is to a pure state. Off-diagonal order measures the
coherences between neighboring sites as the expectation value of the
operator $ \sum_j \sigma_-^{(j)} \sigma_+^{(j+1)}$ evaluated within the
$m_0$-excitation subspace. This parameter emphasizes the effect of the
competing Hamiltonian dynamics as it changes the sign from positive to negative
after the application of a Hamiltonian map, as shown in
figure~\ref{fig:all_comp}. 

Figure~\ref{fig:single_boson} shows an
alternative dataset which demonstrates that the coherent competing Hamiltonian dynamical map
in a system of three sites has - as physically expected - no effect if only a single excitation is
present in the system. Comparison with the data of Fig.~4a in the main text, underlies the significance of the decrease in the overlap with the overlap with the target Dicke state $\ket{D(2,3)}$, and this effect is indeed caused by the competing Hamiltonian dynamics.

Figure~\ref{fig:weak_comp} shows a dataset with competition
and two spin excitations present in the system. Compared to the analysis in the main
text, the competition strength is now set to $\phi = \pi/4$ instead of $\pi/2$. As
expected, this leads to a reduced effect of the competing Hamiltonian maps on the dissipatively created order. 

\renewcommand{\refname}{References}

\end{document}